\documentclass[prd,nofootinbib,showpacs,preprint]{revtex4}
\usepackage{graphicx}
\usepackage{epsfig}
\usepackage{amsmath}
\usepackage{amsfonts}
\usepackage{amssymb}
\usepackage{url}
\usepackage{hyperref}
\usepackage{subfigure}
\usepackage{cancel}
\newcommand{\bd}{\begin{document}}
\newcommand{\ed}{\end{document}}
\newcommand{\bc}{\begin{center}}
\newcommand{\ec}{\end{center}}
\newcommand{\beqa}{\begin{eqnarray}} 
\newcommand{\eeqa}{\end{eqnarray}} 
\newcommand{\beq}{\begin{equation}} 
\newcommand{\eeq}{\end{equation}} 
\newcommand{\lsim}{\lesssim}
\newcommand{\gsim}{\gtrsim}
\newcommand{\nn}{\nonumber}
\newcommand{\bmt}{\begin{pmatrix}}
\newcommand{\emt}{\end{pmatrix}}
\usepackage[toc,page]{appendix}
\usepackage{comment} 
\def\TeV{\mbox{TeV}}
\def\GeV{\mbox{GeV}}
\def\MeV{\mbox{MeV}}
\def\eV{\mbox{eV}}
\def\keV{\mbox{keV}}
\def\s1{\hat s}
\def\ds{\displaystyle}
\def\lb{\Lambda_b}
\def\ll{\Lambda}
\newcommand{\be}{\begin{equation}}
\newcommand{\ee}{\end{equation}}
\newcommand{\bea}{\begin{eqnarray}}
\newcommand{\eea}{\end{eqnarray}}
\newcommand{\bref}[1]{(\ref{#1})}
\def\slashi#1{\rlap{\sl/}#1}
\newcommand{\<}[1]{\langle {#1} \rangle}

\usepackage{verbatim}   
\usepackage{multirow}   

\allowdisplaybreaks[1]

\def \ket#1{ \left| #1 \right\rangle}
\def \brac#1{\left\langle #1 \right|}
\def \abs#1{ \left| #1 \right| }
\def \order#1{ {\cal O} \left( #1 \right) }
\def \dslash#1{ #1\!\!\!/}

\def \eps{\varepsilon}

\def \BtoKll{{\bar{B} \to K \bar{l}l}}
\def \gS{g_s}              
\def \alS{\alpha_s}        
\def \alE{\alpha_e}        
\def \GF{G_F}              
\def \LamConf{{\Lambda_{\rm QCD}}}    
\def \LamConfS{{\Lambda^2_{\rm QCD}}} 

\def \cA{{\cal A}}
\def \cC{{\cal C}}
\def \cF{{\cal F}}
\def \cL{{\cal L}}
\def \cM{{\cal M}}

\def \Gaml{{\Gamma_l}}
\def \Game{{\Gamma_e}}
\def \Gammu{{\Gamma_\mu}}

\def \BRl{{{\cal B}_l}}
\def \BRe{{{\cal B}_e}}
\def \BRmu{{{\cal B}_\mu}}

\def \AFBl{{A_{\rm FB}^l}}
\def \AFBmu{{A_{\rm FB}^\mu}}
\def \AFBe{{A_{\rm FB}^e}}

\def \FHl{{F_H^l}}
\def \FHe{{F_H^e}}
\def \FHmu{{F_H^\mu}}
\begin{document}
\title{New Physics effects in charm meson decays involving $c \to u l^+ l^- (l_i^\mp l_j^\pm)$ transitions }
\author{Suchismita Sahoo and Rukmani Mohanta}
\affiliation{\,School of Physics, University of Hyderabad, 
              Hyderabad - 500046, India  }      
\begin{abstract}
We study the effect of scalar leptoquark and $Z^\prime$ boson on the rare decays of $D$ mesons involving flavour changing transitions $c \to u l^+ l^- (l^\mp_i l^\pm_j)$. We constrain the new physics parameter space using the  branching ratio of the rare decay mode $D^0 \to \mu^+ \mu^-$  and  the  $D^0-\bar D^0$ oscillation data. We compute the branching ratios, forward-backward asymmetry parameters and flat terms in $D^{+(0)} \to \pi^{+(0)} \mu^+ \mu^-$ processes using the constrained parameters.   The branching ratios of the  lepton flavour violating $D$ meson decays, such as $D^0 \to \mu e, ~\tau e$ and $D^{+(0)} \to \pi^{+(0)} \mu^- e^+$ are also investigated.

\end{abstract}
\maketitle
\section{introduction}
The rare $B$ and $D$ meson decay processes driven by flavour changing neutral current (FCNC)  transitions constitute a subject of great interest in the area of electroweak interactions and provide an excellent testing ground to look for new physics beyond the standard model (SM). The FCNC decays are highly suppressed in the SM  and occur  only  at one-loop level. Of particular interest among the FCNC decays are the
rare semileptonic $B$ meson decays involving the   transitions
$b \to s l^+ l^-$, where  several anomalies
at the level of few sigma have been observed recently in the LHCb experiment \cite{lhcb10}.
To compliment these results, efforts should also be made towards the  search for new physics signal  in the up-quark sector, mainly in the  
  rare charm meson  decays involving   $c \to u l^+ l^-$ quark level transitions.
Recently LHCb experiment has  searched for the branching ratio of  the lepton flavour violating (LFV) $D^0 \to \mu^\mp e^\pm$ decays   and put the  limit  as $\rm{BR}(D^0 \to \mu^\mp e^\pm)~ \textless~ 1.3 \times 10^{-8}$ \cite{D0exp}  at $90 \%$ confidence level (CL). On the other hand, both Belle and BaBar experiments have
reported significant deviations on the measured branching fractions of $\bar B \to D^{(*)} \tau \nu_{\tau}$ processes from the corresponding SM predictions. The ratio of these branching fractions, the so-called
$R(D^{(*)})$,  defined as  $R(D^{(*)}) = {\rm BR}(\bar B \to D^{(*)} \tau \nu_{\tau})/ {\rm BR}(\bar B \to D^{(*)} l \nu_l)$, where $l=e, \mu$, exceed the SM prediction by $3.5 \sigma$  \cite{RDstar}, thus open an excellent window to search for new physics (NP) in the up-quark sector.

Mixing between a neutral meson and its anti-meson with a specific flavour  provides an useful tool to deal with  problems in flavour sector. For example, in the past the  $K^0-\bar{K}^0$ and $B^0-\bar{B}^0$  oscillations, involving mesons made up of  down type quarks, had provided information about the charm and top quark mass scale, much before the discovery of these particles in the collider. On the other hand,  $D^0-\bar D^0$ system involves mesons with up-type quarks and  in the SM
the mixing rate is sufficiently small, so that the new physics 
component might play an important role in this case.
The  mixing parameters required to  describe the $D^0-\bar D^0$ mixing are defined by   $x = \Delta M/\Gamma$  and $y= \Delta \Gamma/ 2 \Gamma$, where $\Delta M$
 ($\Delta \Gamma$) is the mass (width) difference between the mass eigenstates. 

In this paper, we focus on the analysis of rare charm meson decays induced by $c \to u \mu^+ \mu^-$ and $c \to u \mu^\mp e^\pm$ FCNC transitions.   We calculate the branching ratios, forward-backward asymmetry parameters and the flat terms in  $D^{+(0)} \to \pi^{+(0)}  \mu^+ \mu^-$  processes both in   the  scalar leptoquark (LQ) and generic $Z^\prime$  model. These processes
suffer from resonance background through $c \to  u M  \to u l^+ l^-$, where $M$ denotes $\eta^{(')}$ (pseudoscalar), $\rho, \phi, \omega$ (vector) mesons. However, to reduce the background coming from these resonances, we work in the low and high $q^2$ regimes, i.e., $ q^2 \in [0.0625, 0.275]{\rm GeV}^2$ and
$ q^2 \in [1.56, 4.00]{\rm GeV}^2$, which lie outside the mass square range of the resonant  mesons. We also compute the  branching ratios of lepton flavour violating $D^0 \to \mu e,  \tau e$ and $D^{+(0)} \to \pi^{+(0)} \mu^- e^+$ processes. These LFV processes have negligible contributions from SM, as they 
proceed through the box diagrams with  tiny neutrino masses in the loop.    However, they can occur at tree level  in the LQ  and $Z^\prime$  models and  are expected to have significantly large branching ratios.
Leptoquarks are  hypothetical color triplet bosonic particles, which couple to quarks and leptons simultaneously and contain both baryon and lepton quantum numbers.  It is  interesting to study  flavour physics with leptoquarks as they allow quark-lepton transitions at tree level, thus explain several  observed anomalies, e.g.,  lepton non-universality (LNU) parameter $R_K|_{q^2 \in[1,6]~{\rm GeV}^2}={\rm BR}(B \to K \mu^+ \mu^-)/{\rm BR}(B \to K e^+ e^-)$ in rare $B$ decays. The existence of scalar leptoquark is predicted in the extended SM  theories, such as grand unified theory  \cite{georgi, georgi2}, Pati-Salam model,  extended technicolor model \cite{schrempp} and the composite model \cite{kaplan}. In this work, we consider the  model which conserves baryon and lepton numbers
and  does not allow proton decay. Here we would like to see how this model affects  the leptonic and semieptonic decays of $D^0$ meson induced by $c \to u l^+ l^-$ transitions.  The phenomenology of scalar leptoquarks and their implications  to the $B$ and $D$-sector  has been  extensively studied in  the literature \cite{Arnold, mohanta1, davidson,leptoquark, Hiller, kosnik, kosnik2, D0-D0bar, ansatz1, Hiller3}.

The $Z^\prime$ boson is a  color singlet  vector gauge boson  and electrically neutral in nature. By adding an additional $U(1)^\prime$ gauge symmetry, the new $Z^\prime$ gauge boson could be naturally derived from the extension of  electroweak symmetry of the SM,  such as  superstring theories, grand unified theories and theories with large extra dimensions.  The processes mediated via   $c \to u$ FCNC transitions could be induced by generic $Z^\prime$ model at tree level.  The theoretical framework of the  heavy new  $Z^\prime$ gauge boson  has been studied in the literature \cite{Pakvasa, Pakvasa2, Arhrib}.  In this paper, we  investigate the $Z^\prime$ contribution to the rare $D^0$ meson decay processes within the parameter space constrained by $D^0-\bar D^0$ mixing and $D^0 \to \mu^+ \mu^-$ processes.

The paper is organized as follows.  In section II, we discuss the   effective Hamiltonian describing  $\Delta C=1$ transitions i.e., $c \to u l^+ l^-$, and  $\Delta C=2$ transition, which is responsible for  $D^0-\bar{D}^0$ mixing. The new physics contribution to  $c \to u$ transitions and the constraint on leptoquark couplings from  $D^0-\bar D^0$ oscillation and $D^0 \to \mu^+ \mu^-$ process  are discussed in section III.  We  calculate the constraint on $Z^\prime$ couplings from $D^0-\bar D^0$ mixing and leptonic $D^0 \to \mu^+ \mu^-$ decays in section IV. In section V, we compute the branching ratios, forward-backward asymmetry parameters and the flat terms of $D^{+(0)} \to \pi^{+(0)} \mu^+ \mu^-$ process in both these models. The  lepton flavour violating $D^{+(0)} \to \pi^{+(0)} \mu^- e^+$  and $D^0 \to \mu e, \tau e $ processes  are discussed in sections VI and VII. Finally we summarize our findings   in section VIII.

\section{Effective Hamiltonian for $\Delta C=1$ and $\Delta C=2$ transitions }
Though the rare charm decays are affected by large non-perturbative effects, the short distance structure of FCNC transitions can be investigated well theoretically.
The change in charm quantum number for rare FCNC charm  meson decays is either of two or one unit, and hence,  involve   either $\Delta C=2$ or $\Delta C=1$ transitions. The $D^0-\bar{D}^0$ mixing takes place via $\Delta C=2$ transition and the decay processes with $\Delta C=1$ transitions are $c \to u l^+ l^-$ and $c \to u \gamma$. 

If we integrate out  the  heavy degrees of freedom associated with the new interactions at a  scale $M$,  an effective Hamiltonian in the form of a series of operators of increasing dimensions can be obtained. However, the operators of dimension $d=6$ have important contributions to charm meson decays or mixing.   In general, one can write the complete basis of these effective operators in terms of chiral quark fields for both $D^0-\bar D^0$ mixing and $D^0 \to l^+ l^-$ process as \cite{Pakvasa, Pakvasa2}
\beq\label{SeriesOfOperators}
\langle f | {\cal H} | i \rangle =
G \sum_{i=1} {\rm C}_i (\mu) ~
\langle f | Q_i  | i \rangle (\mu)  ,
\eeq
where $G$ has inverse-mass squared dimensions, $C_i$ are the Wilson coefficients $\footnote{We denote the Wilson coefficients for $\Delta C =2$ operators as ${c_i}$ and those for $\Delta C =1$ operators as ${C_i}$ through out this work}$.  

The effective operators for $D^0-\bar D^0$ mixing at the heavy mass scale $M$ are given by \cite{Pakvasa, Pakvasa2}
\beqa
\begin{array}{l}
Q_1 = (\overline{u}_L \gamma_\mu c_L) \ (\overline{u}_L \gamma^\mu
c_L)\ , \\
Q_2 = (\overline{u}_L \gamma_\mu c_L) \ (\overline{u}_R \gamma^\mu
c_R)\ , \\
Q_3 = (\overline{u}_L c_R) \ (\overline{u}_R c_L) \ , \\
Q_4 = (\overline{u}_R c_L) \ (\overline{u}_R c_L) \ ,
\end{array}
\qquad
\begin{array}{l}
Q_5 = (\overline{u}_R \sigma_{\mu\nu} c_L) \ ( \overline{u}_R
\sigma^{\mu\nu} c_L)\ , \\
Q_6 = (\overline{u}_R \gamma_\mu c_R) \ (\overline{u}_R \gamma^\mu
c_R)\ , \\
Q_7 = (\overline{u}_L c_R) \ (\overline{u}_L c_R) \ , \\
Q_8 = (\overline{u}_L \sigma_{\mu\nu} c_R) \ (\overline{u}_L
\sigma^{\mu\nu} c_R)\ \ ,
\end{array}
\label{SetOfOperators}
\eeqa
where $q_{L(R)}=L(R)q$ are the chiral quark fields with $L(R)=(1 \mp \gamma_5)/2$ as the projection operators.

In  the standard model, the  effective weak Hamiltonian for $c \to u $ transitions at the scale $\mu=m_c$, can be written as the sum of  three contributions  as \cite{kosnik, faj-ham}
\bea
\mathcal{H}_{eff}=\lambda_d \mathcal{H}^d + \lambda_s \mathcal{H}^s + \lambda_b \mathcal{H}^{\rm peng},
\eea
where $\lambda_q=V_{uq}V_{cq}^*$ is the product of Cabibbo-Kobayashi-Maskawa (CKM) matrix elements.  The explicit form of $\mathcal{H}^{\rm peng}$, which basically responsible for the $c \to u l^+ l^-$ transition  is given by
\bea
\mathcal{H}^{\rm peng}= -\frac{4G_F}{\sqrt{2}}\left (\sum_{i=3,\cdots 10,S,P} C_i \mathcal{O}_i+ \sum_{i=7,\cdots 10,S,P}C_i' \mathcal{O}_i'\right ), \label{SM-Ham}
\eea 
where $G_F$ is the Fermi constant, $C_i$'s are the Wilson coefficients evaluated at the charm quark mass  scale $(\mu=m_c)$ at Next-Next-to-Leading-Order (NNLO) \cite{wilson}. We use the two loop result of Ref. \cite{wilson2} for the  $C_7^{\rm eff}(m_c)$ Wilson coefficients, $V_{cb}^* V_{ub} C_7^{\rm eff} = V_{cs}^* V_{us} (0.007 + 0.020i)(1\pm 0.2)$   and the corresponding  effective operators for $c \to u l^+ l^-$ transitions are given as \cite{kosnik}
\begin{eqnarray}
&&\mathcal{O}_7^{(')} = \frac{e}{16\pi^2}m_c(\bar{u}\sigma_{\mu\nu} R(L) c)F^{\mu\nu}\ , \nn \\
&&\mathcal{O}_9^{(')}=\frac{e^2}{16\pi^2} (\bar{u}\gamma_\mu L(R) c)( \bar{\ell}\gamma^\mu
\ell) \ , \qquad 
\mathcal{O}_{10}^{(')}=\frac{e^2}{16\pi^2} (\bar{u}\gamma_\mu L(R) c)( \bar{\ell}\gamma^\mu
\gamma_5\ell), \nn \\
&&\mathcal{O}_S^{(')}=\frac{e^2}{16\pi^2} (\bar{u} R(L) c)( \bar{\ell}
\ell) \ , \qquad 
~~~~~~\mathcal{O}_{P}^{(')}=\frac{e^2}{16\pi^2} (\bar{u} R(L) c)( \bar{\ell}
\gamma_5\ell),\nn\\
&&\mathcal{O}_T=\frac{e^2}{16\pi^2} (\bar{u} \sigma_{\mu \nu} c)( \bar{\ell} \sigma^{\mu \nu}
\ell) \ , \qquad 
~~~~~\mathcal{O}_{T_5}=\frac{e^2}{16\pi^2} (\bar{u} \sigma_{\mu \nu} c)( \bar{\ell} \sigma^{\mu \nu}
\gamma_5\ell). 
\label{qcdpen}
\end{eqnarray}
The  contributions from the primed operators  as well as    the scalar,   pseudoscalar and tensor operators are absent in the SM and arise only in  beyond  the standard model scenarios. The renormalization group running does not affect the $\mathcal{O}_{10}$ operator, i.e., $C_{10}(m_c) = C_{10}(M_W) \propto  (m_{d,s}^2/m_W^2)$ and hence, the  Wilson coefficient $C_{10}$ is negligible in the SM. 

\section{New physics contribution due to the  exchange of  scalar leptoquarks }
The presence of leptoquarks can modify the SM effective Hamiltonian of $c \to u$ transitions,  giving  appreciable deviation from the SM values. These color triplet bosons can be either scalars or vectors.  There exist three scalar and four vector relevant leptoquark states which potentially contribute to the $c \to u l^+ l^-$ transitions and are invariant under the SM gauge group $SU(3)_C \times SU(2)_L \times U(1)_Y$, where the hypercharge $Y$ is related to the electric charge and weak isospin ($I$)
through $Y= Q-I_3$. Out of  three possible scalar leptoquarks with the quantum numbers $(3,3,-1/3)$, $(3,1,-1/3)$ and $(3,2,7/6)$ \cite{kosnik, kosnik2},  only the leptoquark with multiplet $(3,2,7/6)$  conserves both baryon and lepton numbers and thus, avoids rapid proton decay at the electroweak scale.  Similarly out of the vector multiplets $(3,3,2/3)$, $(3,1,5/3)$, $(3,2,1/6)$ and $(3,2,-5/6)$, only first two leptoquark states 
don't allow baryon number violation  and can be considered to study the observed anomalies in flavour sector.  In this work we consider the baryon number conserving $X=(3,2,7/6)$ scalar leptoquark  which induces the interaction between the up-type quarks and charged leptons and thus, contributes to the semileptonic decay amplitudes.

The interaction Lagrangian  of $X=(3,2,7/6)$ scalar leptoquark   with the SM bilinears  is given by \cite{kosnik, kosnik2}
\bea
\mathcal{L} =\bar{l}_R Y^L \Delta^\dagger Q + \bar{u}_R Y^R \tilde{\Delta}^\dagger L +\rm h.c.,\label{lq-int}
\eea
where $\tilde{\Delta} = i \tau_2 \Delta^*$ represents the conjugate state.  The transition of weak basis to mass basis divides the Yukawa couplings to two part of couplings pertinent for the upper and lower doublet components. The left handed quark and lepton doublets are represented by $Q$ and $L$   and $u_R(l_R)$ is the right handed quark (charged-lepton) singlet. We use the basis where CKM and PMNS rotations are assigned to down type quarks and neutrinos, i.e., $d_L \to V_{CKM} d_L$ and  $\nu_L \to V_{PMNS}   \nu_L$. Here  $Y^{L}$ and $Y^R$ are the leptoquark couplings in the mass basis of the up-type quarks and charged leptons. 
Now writing the leptoquark doublets in terms of its components as $\Delta
=(\Delta^{(5/3)}, \Delta^{(2/3)})^T$, where the superscripts denote the electric charge of the LQ components and expanding the terms in Eqn. (\ref{lq-int}), one can obtain the interaction Lagrangian for different components of LQs 
 given as \cite{kosnik2}
\bea
\mathcal{L}^{(2/3)}&=&\left(\bar{l}_R [Y^L V_{\rm CKM}]  d_L \right) {\Delta^{(2/3)}}^* + \left(\bar{u}_R [Y^R V_{\rm PMNS}] \nu_L \right) \Delta^{(2/3)} + h.c.,\nn\\ 
\mathcal{L}^{(5/3)}&=&\left(\bar{l}_R Y^L   u_L \right) {\Delta^{(5/3)}}^* - \left(\bar{u}_R Y^R l_L \right) \Delta^{(5/3)} + h.c. .\label{int-2}
\eea
Thus, one can see from (\ref{int-2}), that only $\Delta^{(5/3)}$ component mediates the interaction between up-type quarks and charged lepton. Now applying the Fierz transformation, we obtain additional
contributions to the SM Wilson coefficients for $c \to u \mu^+ \mu^-$ transition  as \cite{kosnik}
\bea
&&C_9^{\rm LQ}=C_{10}^{\rm LQ}=-\frac{\pi}{2\sqrt{2}G_F \alpha_{em} \lambda_b} \frac{Y_{\mu c}^L Y_{\mu u}^{L *}}{m_\Delta^2}\ \ , \nonumber\\
&&C_9^{\prime {\rm LQ}}=-C_{10}^{\prime {\rm LQ}}=-\frac{\pi}{2\sqrt{2}G_F \alpha_{em} \lambda_b} \frac{Y_{c \mu }^{R *} Y_{u \mu }^{R}}{m_\Delta^2}\ \ , \nonumber\\
&&C_S^{\rm LQ}=C_{P}^{\rm LQ}=-\frac{\pi}{2 \sqrt{2}G_F \alpha_{em} \lambda_b} \frac{Y_{\mu u}^{L *} Y_{c \mu}^{R *}}{m_\Delta^2}\ \ , \nonumber\\
&&C_S^{\prime {\rm LQ}}=-C_{P}^{\prime {\rm LQ}}=-\frac{\pi}{2\sqrt{2}G_F \alpha_{em} \lambda_b} \frac{Y_{\mu c}^L Y_{u \mu}^{R}}{m_\Delta^2}, \nn \\
&& C_T^{\rm LQ} = -\frac{\pi}{8\sqrt{2}G_F \alpha_{em} \lambda_b} \frac{Y_{u\mu}^R Y_{\mu c}^L + {Y_{c \mu}^R}^* {Y_{\mu u}^L}^*}{m_\Delta^2}, \nn \\
&& C_{T_5}^{\rm LQ} = -\frac{\pi}{8\sqrt{2}G_F \alpha_{em} \lambda_b} \frac{-Y_{u\mu}^R Y_{\mu c}^L + {Y_{c \mu}^R}^* {Y_{\mu u}^L}^*}{m_\Delta^2},
 \label{LQ_Wilson}
\eea
where $\alpha_{em}$ is the fine structure constant. 
After having an idea about the new Wilson coefficients, we now proceed to constrain the combination of LQ couplings  using the  experimental data on $D^0-\bar D^0$ mixing and  $D^0 \to l^+ l^-$ process, where $l=\mu, e$. 

\subsection{Constraint on leptoquark couplings from $D^0 - \bar D^0$ mixing}

In the standard model, $D^0 - \bar D^0$ mixing proceeds through the box diagrams with an internal down-type quarks and $W$-boson exchange and the weak interaction boxes are  suppressed due to GIM mechanism because of the smallness of down-quark mass in comparison to the weak scale. In the LQ model, there will be contribution to the $D^0 - \bar D^0$ mass difference from the box diagrams with the leptoquark  and  leptons  
flowing in the loop.  Since the SM contribution to mass difference is very small, we consider its value to be saturated by new physics contributions. 
Furthermore,  the couplings to the left handed quarks are considered to be zero in order to avoid strict constraints in the down type quark sector. 
Thus, considering only right handed couplings, one can write 
the effective Hamiltonian due to the leptoquark $X(3,2,7/6)$ and charged lepton/neutrinos in the loop as \cite{D0-D0bar}
\begin{eqnarray}
{\cal H}_{eff}=\sum_{l} \frac{(Y_{lc}^{R} Y_{lu}^{R*})^{2}}{128 \pi^2}\left [ \frac{1}{ M_{\Delta}^2}~I
\left (\frac{m_l^2}{M_\Delta^2}\right ) +\frac{1}{M_\Delta^2}\right ](\bar c \gamma^\mu P_R u) (\bar c \gamma_\mu P_R u)\;,
\end{eqnarray}
where the first term is due the charged lepton and second term is due to neutrinos in the loop (ignoring the effect of neutrino mixing). The  loop function  $I(x)$ is given as 
\begin{eqnarray}
I(x)=\frac{1-x^2+2x \log x}{(1-x)^2},
\end{eqnarray}
which is  very close to 1, i.e.,  $I(0)=1$. Using the relation
 \begin{eqnarray}
 \langle \bar D^0|(\bar c
\gamma^\mu P_R u)(\bar c \gamma_\mu P_R u)|D^0 \rangle = \frac{2}{3}  B_D f_{D}^2
M_{D}^2\;,
\end{eqnarray} 
 we obtain the contribution  due to leptoquark exchange as
\begin{eqnarray}
M_{12}^{LQ} =  \frac{1}{2 M_D} \langle \bar D^0 | {\cal H}_{eff}|D^0 \rangle =\frac{\sum_l(Y_{lc }^{R} {Y_{lu}^{R*})^2}}{192 \pi^2 M_\Delta^2}   B_D f_{D}^2 M_{D}\;
\end{eqnarray}
Since $\Delta M_D = 2 |M_{12}|$, we get 
\begin{eqnarray}
\Delta M_{D} = 2 |M_{12}^{LQ}|=  \frac{2}{3} M_D f_D^2 B_D \frac{\left | \sum_l Y_{l c}^{R} {Y_{l u}^{R}}^* \right  |^2}{64 \pi ^2 M_\Delta ^2}\;,
\end{eqnarray}
where $l$ denotes  the charged lepton flavours. In our analysis,   the mass of $D^0$ meson is taken from \cite{pdg}, the value of the  decay constant $f_D=222.6 \pm 16.7^{+2.3}_{-2.4}$ MeV \cite{cleo} and  $B_D (3~ {\rm GeV}) = 0.757(27)(4)$ \cite{BD}. 
To obtain the bound on the leptoquark coupling, we assume that individual leptoquark  contribution to the
mass difference  does not exceed the $1\sigma$ range of the experimental value. Since we are interested to obtain the bounds on 
$Y_{\mu c}^{R} Y_{\mu u}^{R*}$ couplings, here we assume that leptoquark has dominant coupling to muons and its coupling to electron or tau is negligible. 
The SM contribution to the mass difference is very small and hence can be neglected. The corresponding  experimental value  is given by \cite{pdg}
\bea
\Delta M_{ D} = 0.0095^{+ 0.0041}_{-0.0044}~{\rm ps}^{-1}.
\eea

  Now comparing the mass difference with the $1\sigma$ range of  experimental data, the bound on leptoquark coupling for a TeV scale LQ is given by
\bea
7.73 \times 10^{-3} \Big ( \frac{M_\Delta}{1~ {\rm TeV}} \Big ) \leq |Y_{\mu c }^{R} {Y_{\mu u}^{R}}^*| \leq 1.26 \times 10^{-2} \Big ( \frac{M_\Delta}{1~ {\rm TeV}} \Big ),
\eea
which can be translated  with Eqn. (\ref{LQ_Wilson}) to give the constraint  on new Wilson coefficients as
\bea
0.1 ~\Big ( \frac{M_\Delta}{1~ {\rm TeV}} \Big ) \leq \lambda_b 
{ C}_9^{\prime LQ} =-\lambda_b {C}_{10}^{\prime LQ} \leq 0.17~ \Big ( \frac{M_\Delta}{1 ~{\rm TeV}} \Big ).
\eea

\subsection{Constraint  from $D^0 \to \mu^+ \mu^- (e^+ e^-)$ process}
The rare leptonic $D^0 \to \mu^+ \mu^- (e^+ e^-)$ processes, mediated by  FCNC transitions  $c \to u l^+ l^-$ at the quark level,   are highly suppressed in  SM due to negligible $C_{10}$ Wilson coefficient and also suffer from CKM suppression. These processes occur only at one-loop level and are considered as some of the most powerful channels to  constrain the new physics parameter space in the charm-sector. 
Analogous to the  leptonic $B$ meson decay processes, the only non-perturbative quantity involved is the  decay constant of $D$ meson,  which can be reliably calculated using non-perturbative methods such as QCD sum rules, lattice gauge theory and so on. 
 The branching ratio of $D^0 \to l^+ l^-$ process  is given by \cite{Hiller, kosnik}
\bea
{\rm BR}\left(D^0 \to l^+ l^- \right)&=&\tau_D \frac{G_F^2 \alpha_{em}^2 M_D^5 f_D^2 |\lambda_b |^2}{64 \pi^3} \sqrt{1-\frac{4m_l^2}{M_D^2}}  \Bigg[ \Big( 1-\frac{4m_l^2}{M_D^2} \Big)\Big | \frac{C_S^{\rm LQ}-C_S^{\prime {\rm LQ}}}{m_c} \Big |^2 \nn \\ && + \Big | \frac{C_P^{\rm LQ}-C_P^{\prime {\rm LQ}}}{m_c} + \frac{2m_l}{M_D^2} \left(C_{10}^{\rm LQ}-C_{10}^{\prime {\rm LQ}} \right) \Big |^2 \Bigg]. \label{BR_D0}
\eea
The $D^0 \to \mu^+ \mu^-$ process has dominant intermediate $\gamma^* \gamma^*$ state in the SM, which is electromagnetically converted to a $\mu^+ \mu^-$ pair.  After including the contribution of $\gamma^* \gamma^*$ intermediate state, the predicted branching ratio of this process  is  
${\rm BR}(D^0 \to \mu^+ \mu^-)\simeq 2.7 \times 10^{-5} \times{\rm BR}(D^0 \to \gamma \gamma)$ \cite{D0-gamma}.  
 Using the upper bound ${\rm BR}(D^0 \to \gamma \gamma) < 2.2 \times 10^{-6}$ at $90\%$ CL reported in \cite{BaBar}, the estimated limit on branching ratio is   ${\rm BR} \left( D^0 \to \mu^+ \mu^- \right)^{SM} \lesssim 10^{-10}$ \cite{kosnik}.
The  present experimental limits on the branching ratios of dileptonic decays of $D$ meson are  \cite{pdg}
\bea
{\rm BR} \left( D^0 \to \mu^+ \mu^- \right) < 6.2 \times 10^{-9}, ~~~~~~~~~~~~{\rm BR} \left( D^0 \to e^+ e^- \right) < 7.9 \times 10^{-8}.
\eea 
Using the above experimental bounds, the constraint on leptoquark coupling can be obtained by imposing the condition that   individual
leptoquark contribution to the branching ratio does not exceed the experimental limit. 
In this analysis, we neglect the new physics contribution to the 
Wilson coefficient $C_{10}$,  as the scalar and pseudoscalar Wilson coefficients will be dominating due to the additional multiplication factor $M_D/m_l$ as noted from Eqn. (\ref{BR_D0}). 
 Now, redefining the Wilson coefficients as
 \bea \tilde{C}_i^{(')LQ}=\lambda_b C_i^{(')LQ}, \eea we show in Fig. 1, the allowed region  in $\tilde{C}_S^{LQ}-\tilde{C}_S^{\prime LQ}$,  $\tilde{C}_S^{LQ}+\tilde{C}_S^{\prime LQ}$ plane,  obtained from $D^0 \to \mu^+ \mu^-$  (left panel) and $D^0 \to e^+ e^-$ processes (right panel). Here  we have used the relations $C_S^{LQ}=C_P^{LQ}$ and $C_S^{'LQ}=-C_P^{'LQ}$ from Eqn. (\ref{LQ_Wilson}). From the figure, we found the allowed range for the above combinations of Wilson coefficients from $D^0 \to \mu^+ \mu^-$ process as
\bea
\Big| \tilde{C}_S^{LQ}-\tilde{C}_S^{\prime LQ} \Big |~ \leqslant ~ 0.06, ~~~~\Big | \tilde{C}_S^{LQ}+\tilde{C}_S^{\prime LQ} \Big |~ \leqslant ~ 0.06,\label{D0s-LQ}
\eea 
whereas the bounds obtained from $D^0 \to e^+ e^-$ process is rather weak, i.e., 
\bea
\Big| \tilde{C}_S^{LQ}-\tilde{C}_S^{\prime LQ} \Big |~ \leqslant ~ 0.2, ~~~~\Big | \tilde{C}_S^{LQ}+\tilde{C}_S^{\prime LQ} \Big |~ \leqslant ~ 0.2.\label{D0s-e-LQ}
\eea 
It is obvious that  the  bounds obtained in Eqns. (\ref{D0s-LQ}) and (\ref{D0s-e-LQ}) could not give us proper information about the bounds on individual $\tilde C_S^{LQ}$ and $\tilde C_S^{\prime LQ}$ coefficients.
Therefore, we consider only one Wilson coefficient at a time to extract the upper bound on individual coefficients.
In Table I, we report the constraint on $\tilde C_{S,P}^{LQ}$ Wilson coefficients  obtained from the experimental bound on the branching fraction of  $D^0 \to \mu^+ \mu^- (e^+ e^-)$ process.  The bounds  on $\tilde C_i^{\prime LQ}$ Wilson coefficients will be same as  those for
$\tilde C_i^{LQ}$.

If we impose chirality on scalar leptoquarks i.e., they couple to either left-handed or right-handed quarks, but not to both then the $C_{S, P}^{(')}$  Wilson coefficients will  vanish and we get only  the   additional contribution of  $C_{9,10}^{(') LQ}$ Wilson coefficients to the SM.  Now comparing the theoretical and experimental branching ratios, the allowed range of $\tilde{C}_{10}^{(\prime) LQ}$ Wilson coefficients are given in Table I.
\begin{figure}[h]
\centering
\includegraphics[width=7cm,height=6cm]{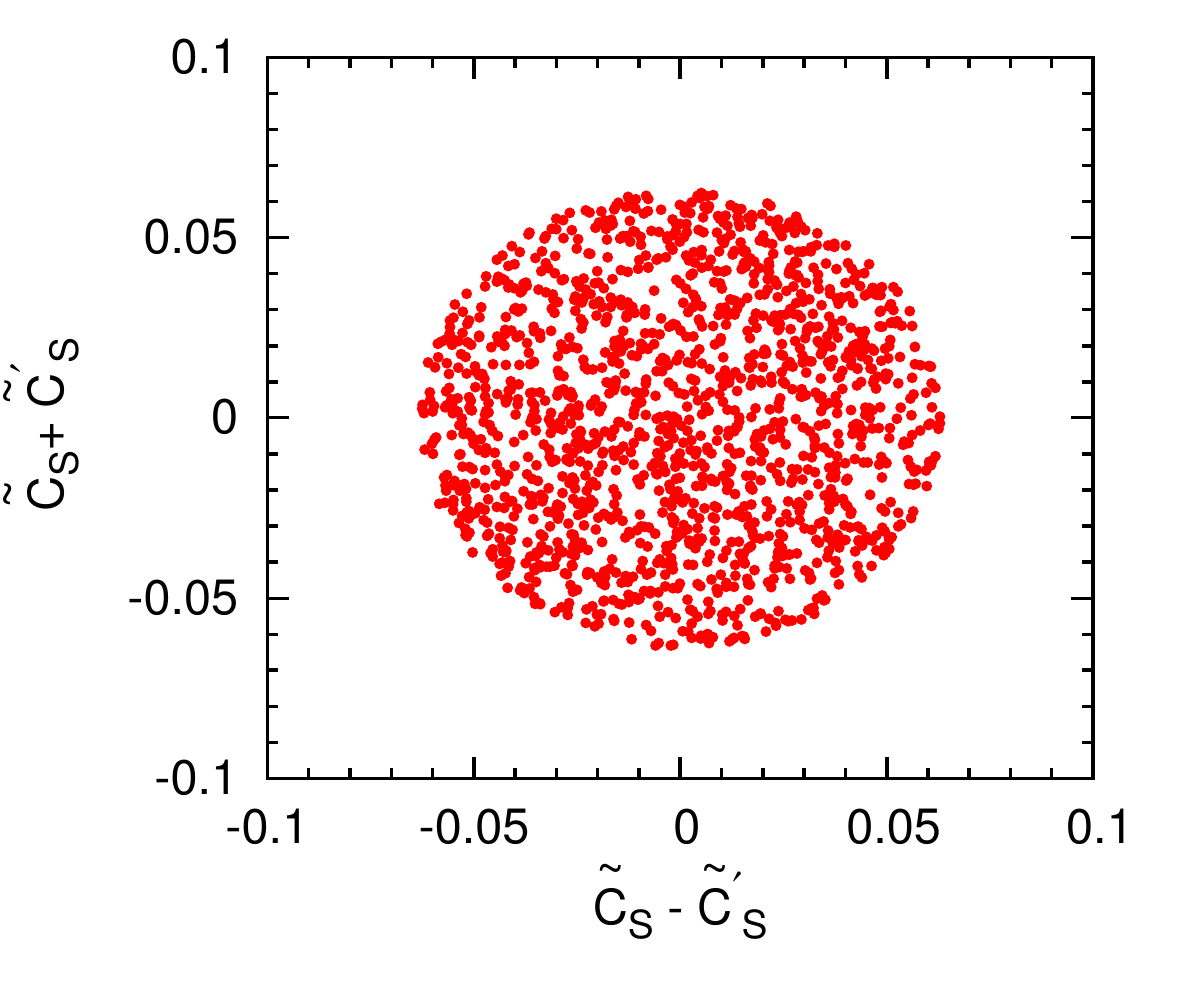}
\includegraphics[width=7cm,height=6cm]{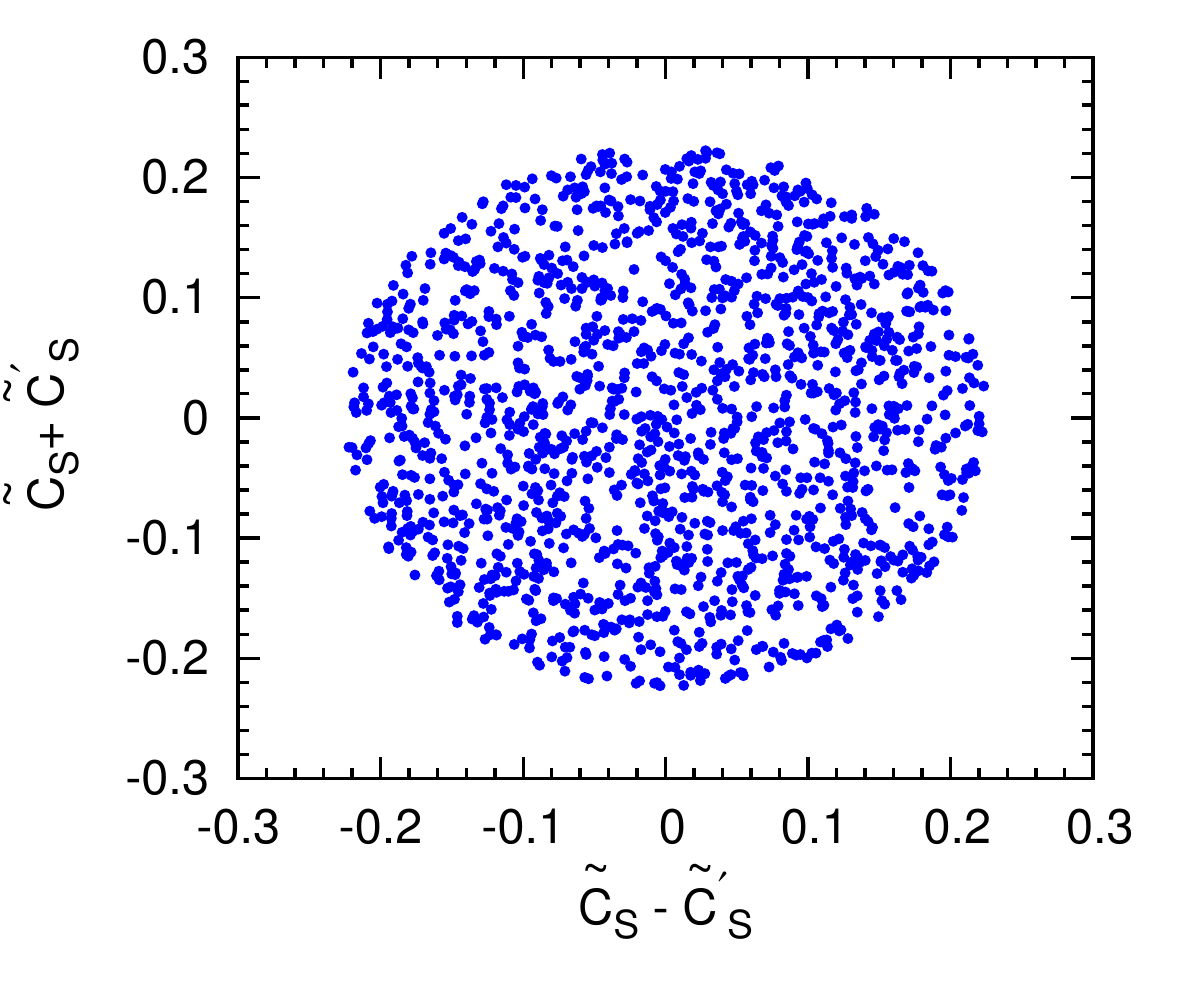}
\caption{The allowed region for $\tilde{C}_S \pm \tilde{C}_S^\prime$ Wilson coefficient obtained from  $D^0 \to \mu^+ \mu^-$  (left panel) and $D^0 \to e^+ e^-$ processes (right panel).}
\end{figure}
\begin{table}[h]
\caption{The allowed values of Wilson coefficients obtained from the upper bound of $D^0 \to \mu^+ \mu^-(e^+ e^-)$ process. The constraint  on $\tilde C_i^{ LQ}$ coefficients can also be applicable to $\tilde C_i^{\prime { LQ}}$ Wilson coefficients.}
\begin{center}
\begin{tabular}{| c | c | c | }
\hline
 Wilson coefficient & $D^0\to  \mu^+ \mu^-$ &$D^0 \to  e^+ e^-$    \\
\hline
$\tilde C_{10}^{ LQ}$  &   $ 0.8$ & $ 600$   \\
$\tilde C_S^{LQ}$  & $0.053$ & $0.186$\\
$\tilde C_P^{LQ}$ & $0.053$   & $0.186$\\
 \hline
\end{tabular}
\end{center}
\end{table}

In order to evade the strict bounds in the down type quark sector, we consider  the leptoquark couplings to the left handed quarks ($Y^L$) as zero. Therefore, the only contribution to the rare charm decays comes   from $\tilde C_9^\prime = -\tilde  C_{10}^\prime$ Wilson coefficients, which are related to  the  right-handed quark couplings.
Now to include   (pseudo)scalar and (pseudo)tensor Wilson coefficients 
and to extract respective  upper bound  complying  with the constraints from $B$ and $K$ physics, we consider a numerically tuned example as discussed in \cite{kosnik}. We assume that the $Y^R$  coupling  is perturbative, i.e., $|Y^R| ~ \textless ~ \sqrt{4\pi}$.  In particular, we consider  a large value for $Y_{c\mu}^R$ coupling, e.g., $Y_{c\mu}^R=3.5$.   We compute the bound on $Y_{u\mu}^R$ coupling by using the constraint  on $\tilde C_{10}^{\prime LQ}$ Wilson coefficients from $D^0 \to \mu^+ \mu^-$  process, which is found to be  small comparatively,   $Y_{u\mu}^R~\textless~8.76 \times 10^{-3}$. Now we instigate  a nonzero coupling to the left handed quark $Y_{u\mu}^L$, which along with the large $Y_{c\mu}^R$ coupling provides nonzero values for $C_{S, P}$ and $C_{T, T_5}$ coefficients. However, the $D^0 \to \mu^+ \mu^-$ process imposes strong bound on $C_S$ coefficient, which together with large $Y_{c\mu}^R$ coupling, limits the left handed coupling as $Y_{\mu u}^L ~\textless ~1.14 \times 10^{-3}$. Thus, from the above discussion  we observe  that  
\bea
\tilde C_9^{LQ} = -\tilde C_{10}^{\prime LQ} = 0.8, ~~~~\tilde C_S^{LQ}=\tilde  C_P^{LQ} = 4\tilde C_T^{LQ} = 4 \tilde C_{T_5}^{LQ} = -0.053.
\eea
Our predicted bound on leptoquark coupling  are in agreement with the constraints obtained in Refs. \cite{kosnik, Hiller} and also with the constraints  obtained from $B$, $K$ physics \cite{carp}.
\section{New physics contribution in  $Z^\prime$ model}
The new heavy $Z^\prime$  gauge boson can  exist in many extended SM scenarios and can mediate the   FCNC  transitions among the fermions in the up-quark sector at  tree level.  The most general Hamiltonian for $c \to u$ transition in the $Z^\prime$ model is given as \cite{Pakvasa2}
\beq
{\cal H}^{\rm FCNC}_{Z'}  = 
{\cal H}^{q}_{Z'} = g_{Z^\prime1} \overline u_L \gamma_\mu c_L Z'^\mu + 
g_{Z^\prime 2} \overline u_R \gamma_\mu c_R Z'^\mu .
\eeq
 Analogously, one can write the Hamiltonian for the  leptonic sector ${\cal H}^{ l}_{Z^\prime}$ as 
\bea\label{VecHamLept}
{\cal H}^L_{Z'} = g_{Z^\prime 1}^\prime \overline \ell_L \gamma_\mu \ell_L Z'^\mu + 
g_{Z^\prime 2}^\prime \overline \ell_R \gamma_\mu \ell_R Z'^\mu.
\eea
Here $g_{Z^\prime i}$ and $g_{Z^\prime i}^\prime$ are the couplings of $Z^\prime$ boson with the quarks and leptons respectively, where $i=1$ or 2 for the  $Z'^\mu$  vector boson coupled to left handed  or right-handed currents.

 After knowing the possible $Z^\prime$ couplings with quarks and leptons, we proceed to constrain the new parameter space  using the results from  charm sector, e.g., the  experimental data on $D^0-\bar{D}^0$ mixing and the branching ratios of $D^0 \to l^+ l^-$ processes. The constraint on the coupling of $Z^\prime$ with the leptonic part  is obtained from the upper limit on branching ratio of lepton flavour violating  $\tau(\mu)^- \to e^- e^+ e^-$ processes.
\subsection{Constraint from $D^0-\bar{D}^0$ mixing}
In this subsection, we calculate the  constraint on $Z^\prime$ couplings from the mass difference of charm meson   mass eigenstates, which  characterizes   the $D^0 - \bar{D}^0$ mixing phenomena.  The $D^0 - \bar{D}^0$ oscillation arises from $|\Delta C =2|$ transition that generates off-diagonal terms in the mass matrix for $D^0$ and $\bar{D}^0$ mesons.
 The mass difference of $D^0 - \bar{D}^0$ mixing at the  scale $\mu = m_c$ is given by \cite{Pakvasa2}
\bea\label{dMV2}
\Delta M_{\rm D}^{\rm (Z^\prime)} = { f_D^2 M_D B_D \over 2 M_{Z^\prime}^2 } 
 \left[ 
{2 \over 3} \left( c_1 (m_c) + c_6 (m_c) \right) - 
\left(\displaystyle{1 \over 2} + \displaystyle{\eta \over 3} \right)
c_2(m_c) + \left( \displaystyle{1 \over 12} + 
\displaystyle{\eta \over 2} \right) c_3(m_c) \right].   \hspace{0.3 cm}   
\eea
At the charm mass scale, the Wilson coefficients in terms of $Z^\prime$ couplings are expressed as
\bea
&&c_1(m_c) = r (m_c,M_{Z^\prime})~ g_{Z^\prime 1}^2\ , ~~~~~~~~~~~~~~
c_3(m_c)= \frac{4}{3} \left[
r(m_c,M_{Z^\prime})^{1/2} - r(m_c,M_{Z^\prime})^{-4}
\right] g_{Z^\prime 1} g_{Z^\prime 2}\ ,\nn\\
&&c_2(m_c)= 2 \ r(m_c,M_{Z^\prime})^{1/2} g_{Z^\prime 1} g_{Z^\prime 2} \ ,~~~~
c_6(m_c)= r (m_c,M_{Z^\prime}) ~ g_{Z^\prime 2}^2  ,
\label{zwilsons}
\eea
where $r (m_c,M_{Z^\prime})$ is the RG factor at the heavy mass scale and  $r (m_c,M_{Z^\prime}) =0.72~(0.71)$ for $Z^\prime$ mass, $M_{Z^\prime} = 1(2)$ TeV \cite{Pakvasa}. 

Now we consider two possible cases to  constrain the couplings   $g_{Z^\prime 1}$ and $g_{Z^\prime 2}$. One with only left handed coupling present, i.e., ($g_{Z^\prime 2} = 0$) and the second where both left handed and right handed couplings are present  with equal strength 
($g_{Z^\prime 1}=g_{Z^\prime 2}=g_{Z^\prime}$).
Here, we  make the simple assumption that the NP part dominates
over the SM contribution in $D^0 - \bar{D}^0$ mixing.
Thus, for the first case, substitution of  $g_{Z^\prime 2}=0$ in  Eqns.   (\ref{dMV2}) and  (\ref{zwilsons}),    the mass difference becomes
\begin{eqnarray}\label{zpr}
\Delta M^{\rm (Z')}_{\rm D} = {f_D^2 M_D B_D r (m_c, M_{Z'})  
\over 3 } ~ {g_{Z'1}^2 \over M_{Z'}^2} \ \ . 
\end{eqnarray}
 Now varying the mass  difference 
$\Delta M_{\rm D}$ within its $1\sigma$ allowed range \cite{pdg}, we obtain
\bea
{g_{Z'1} \over M_{Z'}} = (4.4-7.2) \times 10^{-7}~ {\rm GeV}^{-1},
\eea
and for a representative $Z'$ mass $M_{Z^\prime} =1$ TeV, the value of the coupling is found to be
\bea
g_{Z'1} = (4.4-7.2) \times 10^{-4}.\label{D0-Zprime}
\eea 
Analogously for the second case, i.e.,  $g_{Z^\prime 1}=g_{Z^\prime 2}=g_{Z^\prime}$,  the constraint obtained as  
\bea
g_{Z'}  = (1.2 - 2.0) \times 10^{-4} \Big(\frac{M_{Z^\prime}}{1~ \rm TeV} \Big).
\eea

\subsection{Constraint from $D^0 \to \mu^+ \mu^-$ process}
The effective Hamiltonian for $D^0 \to l^+ l^-$ process in the $Z^\prime$ model is given by \cite{Pakvasa2}
\beq
{\cal H}_{c\to u\ell^+\ell^-}^{Z'} = \frac{1}{M_{Z'}^2} 
\left[
g_{Z^\prime 1}g_{Z^\prime 1}' \widetilde Q_1 +
g_{Z^\prime 1}g_{Z^\prime 2}' \widetilde Q_7 + g_{Z^\prime 1}'g_{Z^\prime 2}  \widetilde Q_2 +
g_{Z^\prime 2}g_{Z^\prime 2}'  \widetilde Q_6 
\right]\ , \label{zprime-Ham}
\eeq where the operators $\widetilde Q_{1,2}$ are 
\bea
\widetilde Q_1 = \left( \bar{l}_L \gamma_\mu l_L \right) \left( \bar{u}_L \gamma^\mu c_L \right), \nn \\
\widetilde Q_2 = \left( \bar{l}_L \gamma_\mu l_L \right) \left( \bar{u}_R \gamma^\mu c_R \right),
\eea
and $\widetilde Q_{6, 7} $ can be obtained from $\widetilde Q_{1, 2} $ by the substitutions of $q_L \to q_R$ and $q_R \to q_L$.

Comparing Eqn. (\ref{zprime-Ham}) with the SM effective Hamiltonian (\ref{SM-Ham}), yields the additional contributions to  Wilson coefficients $C_{9,10}^{(\prime)Z'}$ as
\bea
C_9^{Z^\prime} (C_{10}^{Z^\prime})=-\frac{\pi}{\sqrt{2}G_F \alpha_{em} \lambda_b} \frac{g_{Z^\prime 1} (g_{Z^\prime 2}^\prime \pm g_{Z^\prime 1}^\prime)}{M_{Z^\prime}^2}, \nn \\
{C_9^{\prime}}^{Z^\prime}({C_{10}^{{\prime}^{Z^\prime}}})=-\frac{\pi}{\sqrt{2}G_F \alpha_{em} \lambda_b} \frac{g_{Z^\prime 2} (g_{Z^\prime 2}^\prime \pm g_{Z^\prime 1}^\prime)}{M_{Z^\prime}^2}.
\eea
The branching ratio for $D^0 \to \mu^+ \mu^-$ process in the $Z^\prime$ model is given as \cite{Pakvasa2}
\beq\label{GammaV}
{\rm BR}({D^0 \to \ell^+\ell^-})|_{Z^\prime} = \tau_D\frac{f_D^2 m_\ell^2 M_D}
{32\pi M_{Z^\prime}^4 }
\sqrt{1-\frac{4 m_\ell^2}{M_D^2}} 
\left(g_{Z^\prime 1} - g_{Z^\prime 2}\right)^2 \left(g_{Z^\prime 1}' - g_{Z^\prime 2}'\right)^2\ .
\eeq
For simplicity, we consider here $g_{Z'2}=0$. Now  considering   the couplings of $Z^\prime$ boson to the final leptons as the same form as the  SM-like diagonal couplings of  $Z$ boson to leptons as discussed in  \cite{Pakvasa2}, i.e.,
\beq
g_{Z'1}' = \frac{g}{\cos\theta_W} 
\left(-\frac{1}{2} + \sin^2\theta_W \right), \quad
g_{Z'2}' = g\frac{ \sin^2\theta_W }{\cos\theta_W} \; ,  \label{zprime-lepton-coupling}
\eeq
where $g$ is the gauge coupling of $Z$ boson and $\theta_W$ is the weak mixing angle. Now using the experimental upper limit on branching ratio ${\rm BR}(D^0 \to \mu^+ \mu^-) ~\textless ~6.2 \times 10^{-9}$ \cite{pdg}, we obtain 
\bea
{g_{Z'1} \over M_{Z'}^2} ~\textless ~7.67 \times 10^{-8}~ {\rm GeV}^{-2} .
\eea
 For $M_{Z^\prime}=1$ TeV, the constraint on $g_{Z'1}$ coupling is
\bea 
  g_{Z'1}~ \textless ~0.077,\label{D0mu-Zprime}
\eea  
   which is rather weak compared to the constraint obtained from $D^0-\bar{D}^0$ mixing.

It should be noted that the constraint on $Z^\prime$ couplings from the $D^0 \to \mu^+ \mu^-$ decay  process and the $D^0 - \bar D^0$ mixing data have been computed in  \cite{Pakvasa2}. Similarly, the  constraint on the couplings  from $D^0 - \bar D^0$ oscillation are obtained in Ref. \cite{Pakvasa}.    We found that our constraints are consistent with the above predictions,  if we use the updated values of various input parameters.

\subsection{Constraints on $g_{Z1}'$ from $\tau^-(\mu^-) \to e^- e^+ e^-$ process}
Considering only the left handed coupling to the $Z^\prime$ boson, the branching ratio of $\mu^- \to e^- e^+ e^-$ process in the $Z^\prime$ model is given by \cite{matias, mu-to-3e}
\bea
{\rm BR}\left(\mu^- \to e^- e^+ e^- \right) = \frac{\tau_\mu m_\mu^5}{768 \pi^3} \frac{ |{g'}_{\mu e}^L {g'}_{ee}^L |^2 }{M_{Z^\prime}^4}\;,
\eea
where we have explicitly shown the indices in the couplings.
The experimental  upper limit on branching ratio of this mode is ${\rm BR} (\mu^- \to e^- e^+ e^-) < 10^{-12}$ \cite{pdg}. For the analysis, we use the mass and lifetime of muon from \cite{pdg} and  consider 
the coupling  ${g'}_{ee}^L=g_{Z' 1}^\prime$   as SM-like with its value as presented in  Eqn.(\ref{zprime-lepton-coupling}).  Thus, using the experimental upper limit, we get the  bound  on ${g'}_{\mu e}^L$ coupling  as  
\bea
|{g'}_{\mu e}^L| ~\textless ~5.69 \times 10^{-5} \Big(\frac{M_{Z^\prime}}{1~ \rm TeV} \Big).\label{D0mueee-Zprime}
\eea 
Analogously using the  branching ratio of $\tau^-\to e^- e^+ e^-$ process,
  ${\rm BR} (\tau^- \to e^- e^+ e^-) <  2.7 \times 10^{-8}$  \cite{pdg}, the constraint on lepton flavour violating ${g'}_{\tau e}^L$ coupling is found to be
\bea
|{g'}_{\tau e}^L| ~\textless ~0.02 \Big(\frac{M_{Z^\prime}}{1 ~\rm TeV} \Big). \label{D0taueee-Zprime}
\eea

\section{$D^{+ (0)} \to \pi^{+ (0)} \mu^+ \mu^-$ process}
In this section, we study the rare semileptonic  decay process
$D^+ \to \pi^+ \mu^+ \mu^-$, which is mediated by the quark level transition $c \to u \mu^+ \mu^-$  and constitutes a suitable tool to search for new physics.  The dominant resonance contributions come from the $\phi$, $\rho$ and $\omega$ vector mesons and   the effects of  $\eta^{(')}$ mesons are comparatively  negligible. 
These decay modes are recently studied in Refs. \cite{Hiller, kosnik, rim} in various new physics scenarios and it is found that the model with scalar/vector leptoquarks and minimal supersymmetric model with R-parity violation can give significant contributions.
The matrix elements of  various hadronic currents between
the initial $D$ meson and the final $\pi$ meson  can be parametrized in terms of three form factors $f_0$, $f_+$ and $f_T$    \cite{kosnik} as
\bea
\langle \pi(k)| \bar u \gamma^\mu (1\pm \gamma_5) c | {D}(p_D) \rangle &=&f_+\left(q^2\right) \Big[ \left(p_D+k\right)^\mu -\frac{M^2_D - M^2_\pi}{q^2}q^\mu \Big]\nn\\
&+& f_0\left(q^2\right) \frac{M^2_D - M^2_\pi}{q^2}q^\mu, \hspace{0.3cm}
\eea
\bea
\langle \pi(k)| \bar u \sigma^{\mu \nu} (1\pm \gamma_5) c|{D}(p_D) \rangle & =& i\frac{f_T(q^2)}{M_D+M_\pi}[\left(p_D+k\right)^\mu  q^\nu\nn\\ & -& \left(p_D+k\right)^\nu  q^\mu \pm i \epsilon^{\mu \nu \alpha \beta} \left(p_D+k\right)_\alpha  q_\beta ], \hspace{0.3cm}
\eea
where $p_D$ and $k$ are the four momenta of the $D$  and $\pi$ mesons respectively and $q=p_D-k$ is the momentum transfer. The form factors for $D^0 \to \pi^0$ are scaled as $f_i \to f_i/\sqrt{2}$ by isospin symmetry.  For the $q^2$-dependence of the form factors, we use the parameterization  from  Refs. \cite{formfactor1, formfactor2},  as
\bea
f_+(q^2) = \frac{f_+(0)}{(1-x)(1-ax)}\ , ~~~  f_0(q^2) = \frac{f_+(0)}{(1-(x/b))}\ ,~ ~~ f_T(q^2) = \frac{f_T(0)}{(1-x_T)(1-a_T x_T)}\ ,\hspace{0.3 true cm}
\eea
where $x=q^2/m_{pole}^2$ with $m_{pole}=1.90(8)$ GeV, $a=0.28(14)$ and $b=1.27(17)$  are  the shape parameters \cite{kosnik} measured from $D \to \pi l \nu$ decay process and $f_+(0)=0.67(3)$ \cite{fplus}. The parameters in the $f_T$ form factor are:  $x_T=q^2/M_{D^*}^2$, $f_T(0)=0.46(4)$ and $a_T=0.18(16)$ \cite{formfactor2}. Thus, one can write the transition amplitude for $D^+ \to \pi^+ \mu^+ \mu^-$ process  as \cite{Hiller2, kosnik}
\bea
  \label{eq:matrix:el}
  \cM(D^+ \to \pi^+ l^+ l^-)  &=& i \frac{\GF  \lambda_b \alpha_{em}}{\sqrt{2} \pi}
     \Bigg(V\, p_D^{\mu}\, [\bar{l}\gamma_{\mu} l] + A\, p_D^{\mu}\,
     [\bar{l} \gamma_{\mu}\gamma_5 l] + \left( S + T\cos\theta \right) [\bar{l}l] 
     \nn \\ && + \left(P + T_5 \cos \theta \right ) [\bar{l}\gamma_5 l] \Bigg),
\eea
where $\theta$ is the angle between the  $D$ meson and the negatively charged lepton in the rest frame of the dilepton.   The functions $V$, $A$, $S$ and $P$ are defined in terms of the Wilson coefficients  as 
\bea
V &=&  \frac{2 m_c f_T(q^2)}{M_D + M_\pi} C_7+ f_+(q^2)  (C_9+C_9^{NP}+C_9^{\prime NP}) ,\nn \\ 
A  &=&f_+(q^2) \left(C_{10}+C_{10}^{NP}+C^{\prime NP}_{10}\right),\nn  \\  
S&=& \frac{M_D^2 - M_\pi^2}{2m_c} f_0(q^2) (C_S^{NP} + C^{\prime NP}_S),\nn \\ 
 P & =& \frac{M_D^2-M_\pi^2}{2m_c} f_0(q^2) (C_P^{NP} + C^{\prime NP}_P) \nn \\ 
       & -& m_l   \left[ f_+(q^2)- \frac{M_D^2 - M_\pi^2}{q^2} \left(
         f_0(q^2)-f_+(q^2) \right) \right] \left(C_{10}+C_{10}^{NP}+C^{\prime NP}_{10}\right), \nn \\
T &=& \frac{2f_T(q^2) \beta_l \lambda^{1/2}}{M_D + M_\pi} C_T^{NP}, \nn \\  
T_5 &=& \frac{2f_T(q^2) \beta_l \lambda^{1/2}}{M_D + M_\pi} C_{T_5}^{NP}.       
\eea
Here  $C_{9, 10}^{(')NP}$, $C_{S, P}^{(')NP}$ and $C_{T, T_5}^{NP}$  are the  new Wilson coefficients arising from either the scalar leptoquark model or the generic $Z^\prime$ model.
Using Eqn. (\ref{eq:matrix:el}), the double differential decay distribution  with respect to $q^2$ and $\theta$, for the lepton flavour $l$ is given by \cite{kosnik, Hiller2}
\begin{equation}
  \label{eq:double:dG}
  \frac{d^2\Gaml}{dq^2\, d\!\cos\theta} = 
    a_l(q^2) + b_l(q^2) \cos\theta + c_l(q^2) \cos^2\theta,
\end{equation}
where
\bea
  a_l(q^2)  & = & \Gamma_0\, \sqrt{\lambda}\, \beta_l \Big\{ 
    2q^2 \left( \beta^2_l |S|^2 + |P|^2 \right)
    + \frac{\lambda}{2}  (|A|^2 + |V|^2) \nn  \\ && 
    + 4 m_l (M_D^2 - M_\pi^2 + q^2){\rm Re}(A P^\ast) + 8 m_l^2 M_D^2 |A|^2 \Big\}, \nn \\
 b_l(q^2)  & = & 4  \Gamma_0\, \sqrt{\lambda}\, \beta_l \Big\{ q^2 \beta_l^2 {\rm Re}(S T^\ast) +q^2 {\rm Re} (P T_5^\ast) \nn \\ && + m_l (M_D^2 - M_\pi^2 +q^2) {\rm Re} (A T_5^\ast) +\sqrt{\lambda}\, \beta_l  m_l    {\rm Re}(V S^\ast)\Big\} , \nn \\ 
c_l(q^2) & =  & \Gamma_0\, \sqrt{\lambda}\, \beta_l  \Big\{  - \frac{\lambda \beta_l^2}{2} (|V|^2 + |A|^2) +2q^2 (\beta_l^2 |T|^2 +|T_5|^2) \nn \\ && + 4m_l \beta_l \lambda^{1/2} {\rm Re} (V T^\ast) \Big\},
\eea
with
\begin{align}
\lambda &= M_D^4 + M_\pi^4 + q^4 - 2 \left(M_D^2 M_\pi^2 + M_D^2 q^2 + M_\pi^2 q^2\right), &
  \beta_l & = \sqrt{1 - 4 \frac{m_l^2}{q^2}},
\end{align}
and 
\begin{equation}  \label{eq4}
  \Gamma_0 = \frac{\GF^2 \alpha_{em}^2 |\lambda_b|^2}{ (4\pi)^5 M_D^3}.
\end{equation}
Thus, the branching ratio is given by
\bea
\frac{d{\rm BR}}{dq^2} = 2\tau_D  \Big[a_l(q^2)+\frac{1}{3} c_l(q^2)\Big]. 
\eea
The  forward-backward asymmetry ($A_{FB}$) is another  useful observable to look for new physics, which is defined as \cite{kosnik}
\bea
A_{FB} (q^2) = \Bigg[\int_0^1 d\cos \theta \frac{d^2\Gamma}{dq^2 d \cos\theta}-\int_{-1}^0 d\cos\theta\frac{d^2\Gamma}{dq^2 d \cos\theta}\Bigg] \Bigg  /\frac{d\Gamma}{dq^2} = \frac{b_l(q^2)}{a_l(q^2)+\frac{1}{3}c_l(q^2)}.
\eea
Since the coefficient $b_l$ depends only on scalar and pseudoscalar Wilson coefficients the forward-backward asymmetry is zero in  SM. However, the additional new physics contribution can give non-zero contribution to the forward-backward asymmetry parameter.   Another interesting observable 
is the flat term,   defined as \cite{Hiller2}
\begin{equation}
F^l_H = \int_{q^2_{min}}^{q^2_{max}}dq^2 \left(a_l + c_l\right)\Big / {\int_{q^2_{min}}^{q^2_{max}}dq^2 \left(a_l + \frac{1}{3}c_l\right) },
\end{equation}
where the  uncertainties get reduced due to the cancelation between the numerator and denominator.

For numerical evaluation, we  take the particle masses and the lifetime of $D$ meson from \cite{pdg}.  For the CKM matrix elements, we use the Wolfenstein parametrization with values $A=0.814_{-0.024}^{+0.023}$, $\lambda=0.22537\pm 0.00061$, $\bar{ \rho} =0.117 \pm 0.021$ and $\bar{\eta} =0.353 \pm 0.013$ \cite{pdg}.    With these input parameters,  we compute the resonant/non-resonant  branching ratios of  $D^{+} \to \pi^{+} \mu^+ \mu^-$ process by integrating the decay distribution with respect to $q^2$. 
We parametrize the contributions from the  resonances with the Breit-Wigner shapes for $C_9 \to C_9^{\rm res}$, for $ \rho, \omega, \phi$ (vector)  and $C_P \to C_P^{\rm res}$ for  $\eta^{(\prime)}$ (pseudoscalar)  mesons as \cite{Hiller, kosnik}
\bea
&&C_9^{\rm res} =   a_\rho e^{i\delta_\rho} \Bigg ( \frac{1}{q^2-m_\rho^2 +im_\rho \Gamma_\rho} - \frac{1}{3} \frac{1}{q^2-m_\omega^2 +im_\omega \Gamma_\omega} \Bigg ) + \frac{a_\phi e^{i\delta_\phi}}{q^2-m_\phi^2 +im_\phi \Gamma_\phi}, \nn \\
&&C_P^{\rm res} =    \frac{ a_\eta e^{i\delta_\eta}}{q^2-m_\eta^2 +im_\eta \Gamma_\eta} +  \frac{a_{\eta^\prime} }{q^2-m_{\eta^\prime}^2 +im_{\eta^\prime} \Gamma_{\eta^\prime}}. 
\eea 
Here $m_{M} ~(\Gamma_{M})$ denotes the mass (total decay width) of the resonant state $M$, where $M$ corresponds to $ \eta^{(\prime)}, \rho, \omega, \phi$ mesons. With the approximation of ${\rm BR}(D^+ \to \pi^+ M(\to \mu^+ \mu^-)) \simeq {\rm BR}(D^+ \to \pi^+ M) {\rm BR}(M \to \mu^+ \mu^-)$ and considering the experimental upper bound from \cite{pdg}, the magnitudes of the Breit-Winger parameters are given by \cite{Hiller}
\bea \label{a_M}
&&a_\phi = 0.24^{+0.05}_{-0.06}~{\rm GeV}^2, ~~~~a_\rho  =  0.17 \pm 0.02 ~{\rm GeV}^2,~~~~ a_\omega = a_\rho/ 3, \nn \\
&& a_\eta = 0.00060^{+0.00004}_{-0.00005} ~{\rm GeV}^2,~~~~ a_{\eta}^\prime \sim 0.0007 ~{\rm GeV}^2.   
\eea
   The detailed procedure of SM resonant  contributions to $D^+ \to \pi^+ \mu^+ \mu^-$  process can be found in \cite{formfactor2, kosnik, Hiller}.  In Fig. 2, we show the $q^2$ variation of  branching ratio of $D^+ \to \pi^+ \mu^+ \mu^-$  process  including the resonant contribution in the SM. The band in the figure is  due to the uncertainties associated with the $a_M$ parameters as given in (\ref{a_M}) and the random variation of  relative phases within $-\pi$ and $\pi$. For simplicity we have assumed the same  phase for all the resonances. From the figure, one can observe that in
the low and high $q^2$ regions the long distance resonant contributions are approximately one order of magnitude below the current experimental sensitivity, and hence these  regions are suitable  to look for new physics beyond the SM.    Thus, both  in the SM and in the  leptoquark and $Z^\prime$ models, we study the $D^+ \to \pi^+ \mu^+ \mu^-$ process only at  the very low and high $q^2$ regimes. However, it should be emphasized that  the uncorrelated variation of the unknown resonant phases affects the branching ratio in the low $q^2$ region significantly, which makes it quite difficult to infer the  possible role of new physics.

With all the input parameters from \cite{pdg}  along with  the SM Wilson coefficients \cite{wilson, wilson2},  we present in Table II, the predicted values  of branching ratios for  the $D^{+(0)} \to \pi^{+(0)} \mu^+ \mu^-$  processes by integrating the decay distribution  in low and high $q^2$ bins.  Here we have used the $q^2$ regimes as $q^2 \in [0.0625, 0.275] ~{\rm GeV}^{2}$ and $q^2 \geq 1.56~{\rm GeV}^2$ to reduce the background coming from the dominant resonances. The theoretical uncertainties in the SM are   associated with the lifetime of $D$ meson, CKM matrix
elements and the hadronic form factors. In Fig. 3, the variation of SM  branching ratios  of  $D^{+} \to \pi^{+} \mu^+ \mu^-$ process in the very low and high $q^2$ regimes are shown in red dashed lines and the green bands represent the SM theoretical uncertainties.    
    
Now using the constraint on the leptoquark  parameter space
obtained in section III,  we show in Fig. 3,  the $q^2$ variation of branching ratio of $D^+ \to \pi^+ \mu^+ \mu^-$ process in low $q^2$ (left panel) and high $q^2$  (right panel) both in the  scalar leptoquark
and $Z'$  models. Here the orange (blue) band represents the contributions from the scalar leptoquark ($Z'$) model.  The $90\%$ CL experimental upper bounds on the branching ratios from \cite{LHCb-exp} 
\bea
{\rm BR}(D^+ \to \pi^+ \mu^+ \mu^-)|_{\rm low~q^2} < 2.0 \times 10^{-8},\nn\\
{\rm BR}(D^+ \to \pi^+ \mu^+ \mu^-)|_{\rm high~q^2} < 2.6 \times 10^{-8},
\eea
are shown in thick black lines.   In Table II, we present the integrated branching ratios of $D^{+(0)} \to \pi^{+(0)} \mu^+ \mu^- $ processes  in both the low and high $q^2$ regions in the leptoquark and $Z^\prime$ models.  We find that the predicted branching ratios in the leptoquark model have significant deviations from the corresponding SM values due to the effect of scalar leptoquark and are well below the experimental upper limits. However the effect of $Z^\prime$ boson to the branching ratios of $D^{+(0)} \to \pi^{+(0)} \mu^+ \mu^-$ processes is very marginal.

In the leptoquark model, the variation of forward-backward asymmetry for  $D^+ \to \pi^+ \mu^+ \mu^-$ process in low $q^2$ (left panel) and high $q^2$ (right panel) is presented in Fig. 4.  The forward-backward asymmetry depends on the combinations of  $C_{S}^{(\prime)}$ and $C_{T, T_5}$ Wilson coefficients, thus  have zero value in the SM. However, the additional contributions of $C_{S,P}^{'}$ Wilson coefficients due to scalar leptoquark exchange  give non-zero contribution to the forward-backward asymmetry, though it is not so significant. The integrated forward-backward asymmetry for $D^+ \to \pi^+ \mu^+ \mu^-$ process is given as
\bea
\langle A_{FB} \rangle & =& -0.083 \to 0.042 ~~~{\rm ~in ~ low ~q^2}, \nn \\
\langle A_{FB} \rangle &=& -0.087 \to 0.062 ~~~{\rm ~in ~ high ~q^2}, \nn \\
\langle A_{FB} \rangle &=& -0.095 \to 0.06 ~~~{\rm ~in ~ full ~q^2}.
\eea
 The $Z^\prime$ model provides additional contributions  only to the $C_{9, 10}$ Wilson coefficients, and there are no new contributions to    scalar or tensor  terms. Thus, the forward-backward asymmetry  vanishes in the $Z^\prime$ model.
In both the LQ and $Z^\prime$ model,  the plot for flat term of $D^+ \to \pi^+ \mu^+ \mu^-$ process with respect to low $q^2$ (left panel) and high $q^2$ (right panel) is given in Fig. 5. The predicted values in low $q^2$ range are 
 \bea
 \langle F_H \rangle |_{\rm SM} = 0.4 \pm 0.064, ~~~\langle F_H \rangle |_{\rm LQ}=0.336 \to 0.46, ~~~ \langle F_H \rangle |_{\rm Z^\prime}=0.4 \to 0.41,
\eea 
 and in the region of high $q^2$
 \bea
 \langle F_H \rangle |_{\rm SM} = 0.03 \pm 0.005, ~~~\langle F_H \rangle |_{\rm LQ}=0.34 \to 0.5, ~~~ \langle F_H \rangle |_{\rm Z^\prime}=0.08 \to 0.095.
\eea 

In addition to the leptoquark and $Z^\prime$ models, the rare charm meson decays mediating by the $c \to u$ transitions have also been investigated in various new physics models such as, Minimal Supersymmetric Standard Model \cite{D0-gamma, rim, Fajfer2, Fajfer3},  two Higgs doublet model \cite{Fajfer2}, warped extra dimensions model \cite{Fajfer3} and the up vector like quark singlet model \cite{Delaunay}. 
 In the Ref. \cite{Hiller, kosnik}, the $D \to \pi \mu^+ \mu^-$ process is studied in the context of both scalar and vector leptoquark models. Our predicted results are found to be consistent with the literature.
 
 
\begin{figure}[h]
\centering
\includegraphics[width=7.6 cm,height=5.8cm]{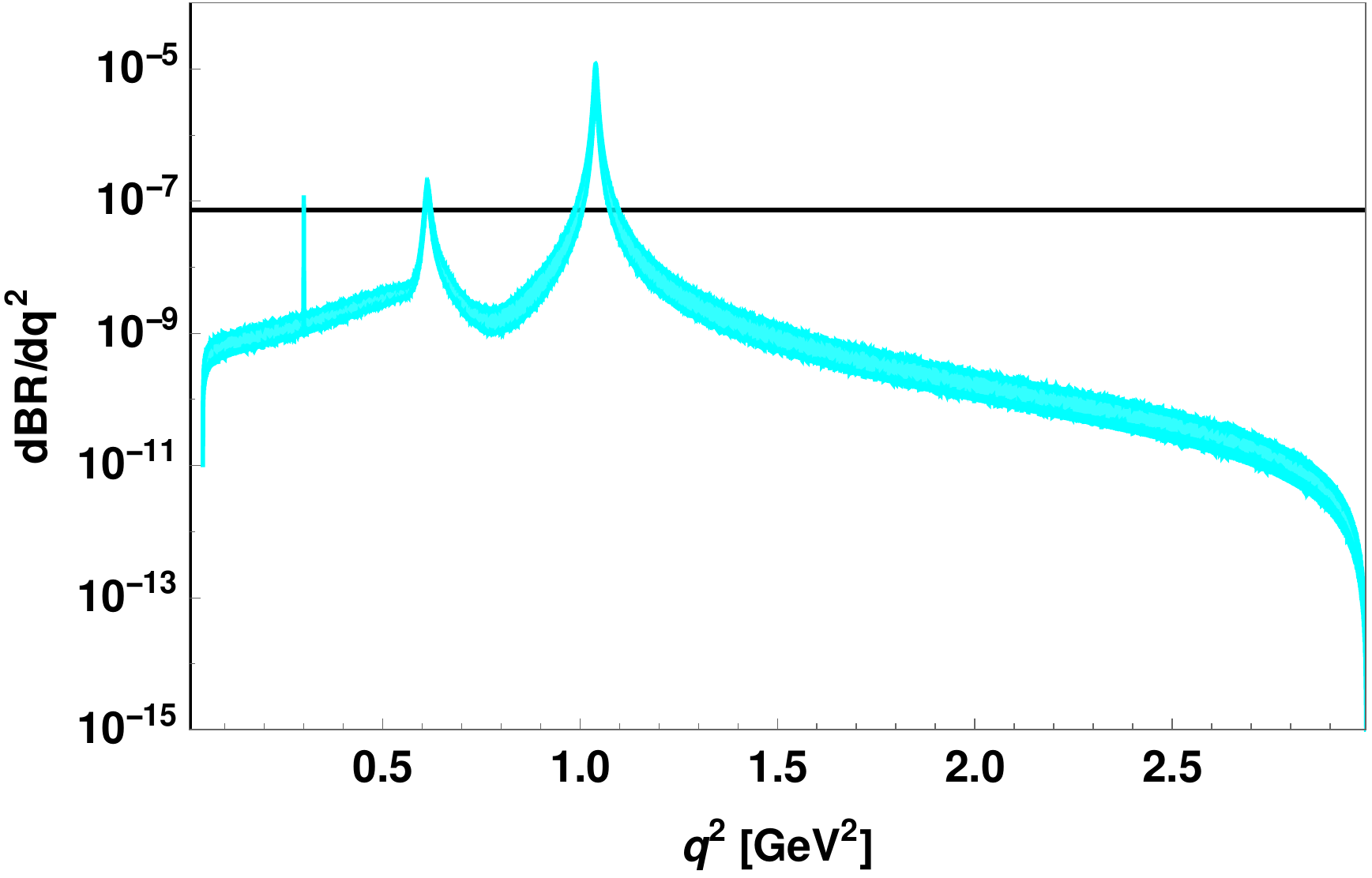}
\caption{The resonant contributions to the  branching ratio of $ D^+ \rightarrow \pi^+ \mu^+ \mu^-$  in the SM.  The band arises due to the uncertainties in  Breit-Winger parameters and the variation of relative phases. The horizontal black line represents the experimental upper bound from \cite{pdg}. }
\end{figure}
\begin{figure}[h]
\centering
\includegraphics[width=7.0cm,height=5.5cm]{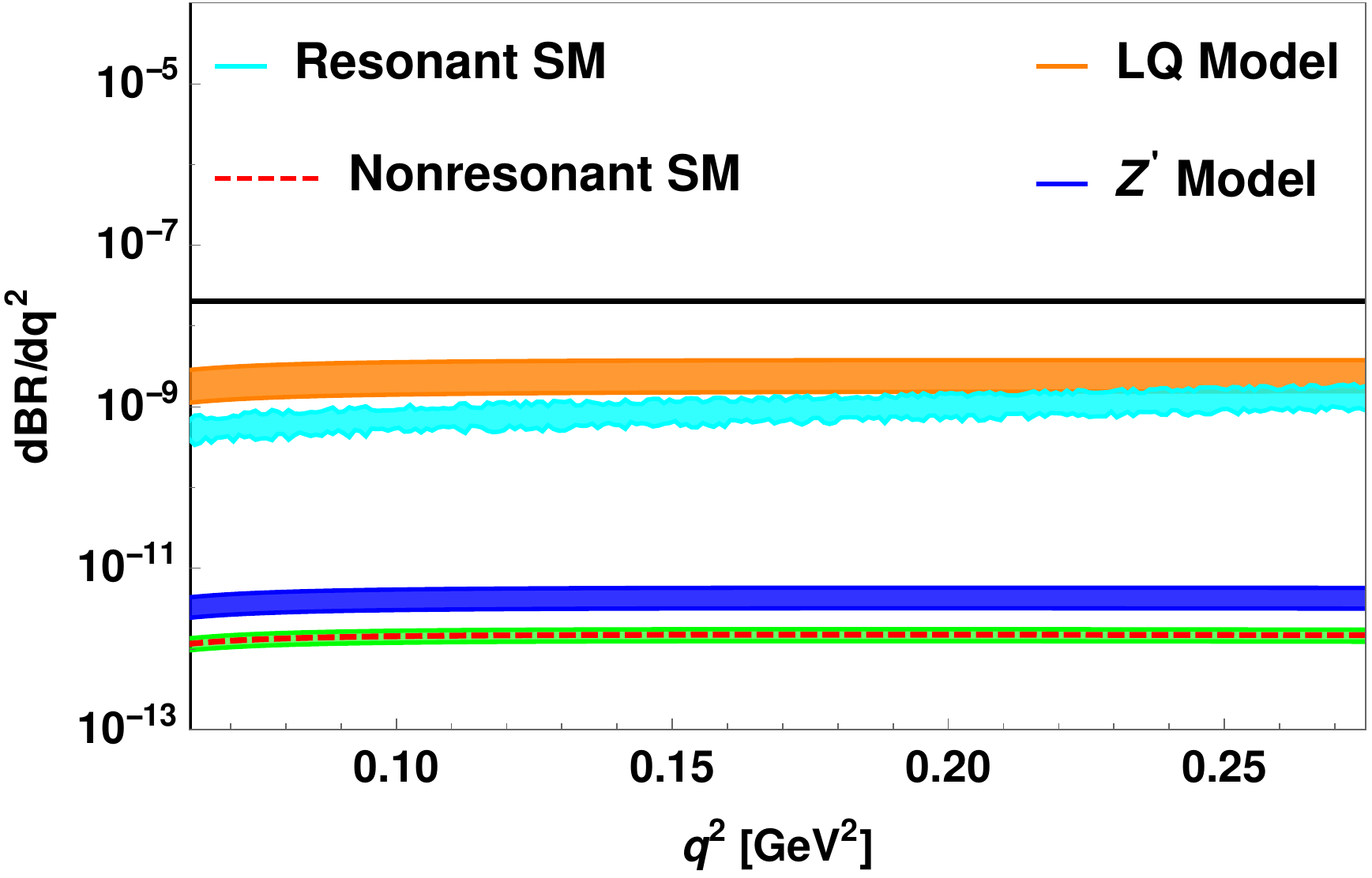}
\hspace{0.25 true cm}
\includegraphics[width=7.0cm,height=5.5cm]{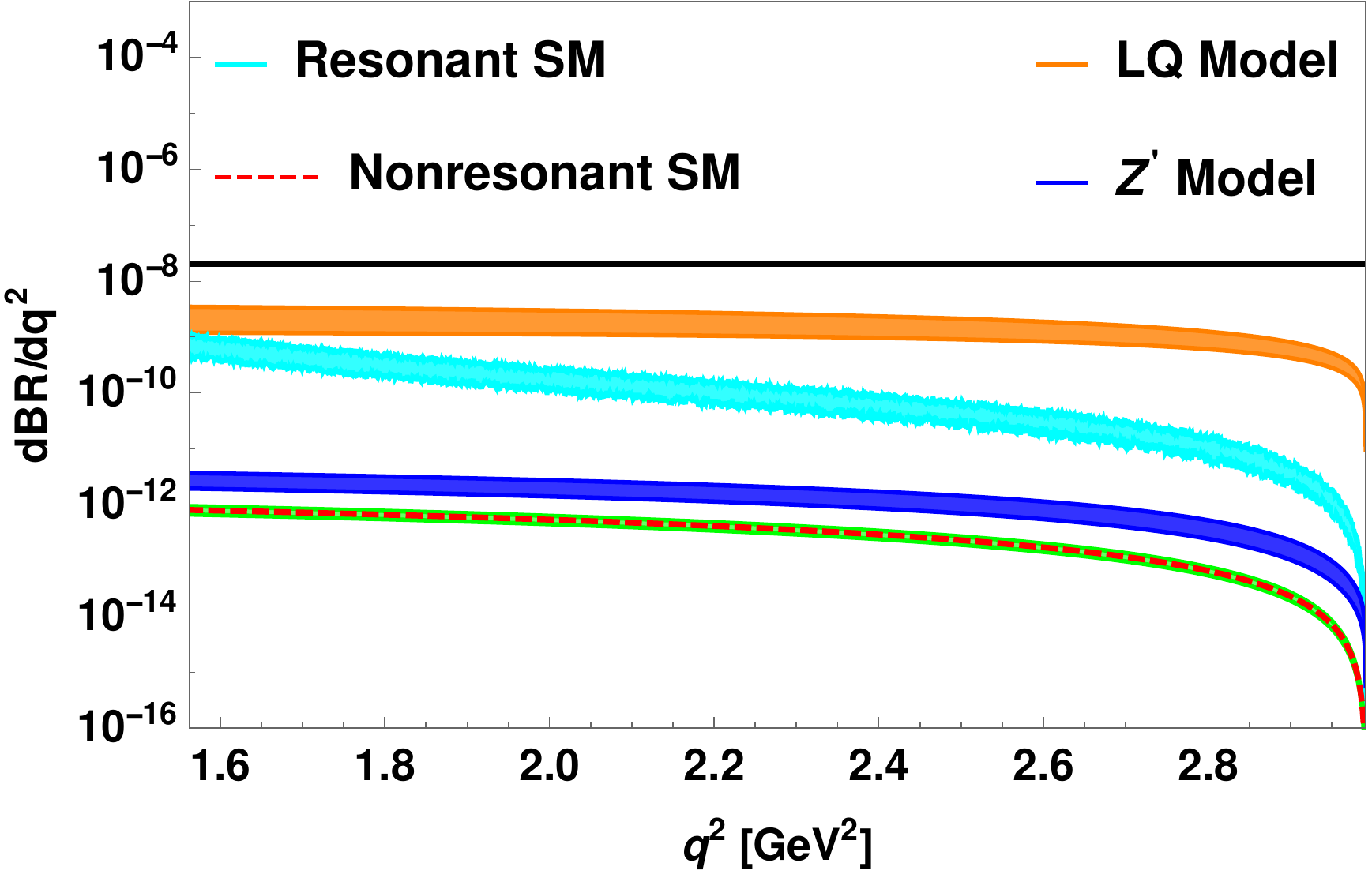}
\caption{The variation of branching ratio of $ D^+ \rightarrow \pi^+ \mu^+ \mu^-$  with respect to low $q^2$ (left panel) and high $q^2$ (right panel). The orange bands represent the contributions from scalar leptoquark, the blue bands are due to   the $Z^\prime$ contributions, the red dashed lines are  for non-resonant SM and the cyan bands are for resonant SM.  The green bands stand for the theoretical uncertainties from the input parameters in the SM.  The solid black line denotes the $90\%$ CL experimental upper limit \cite{LHCb-exp}.}
\end{figure}

\begin{figure}[h]
\centering
\includegraphics[width=7.0cm,height=5.5cm]{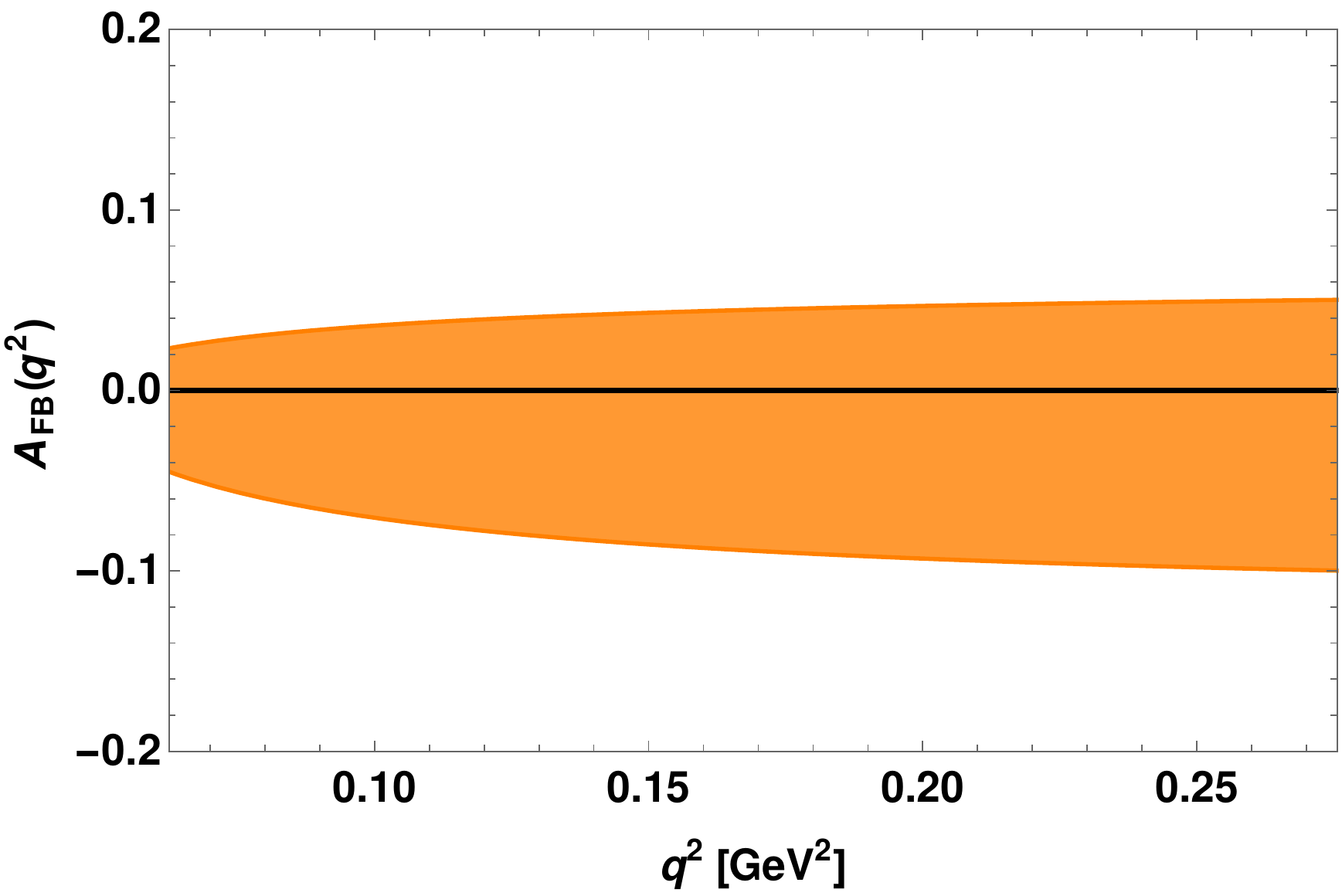}
\hspace{0.25 true cm}
\includegraphics[width=7.0cm,height=5.5cm]{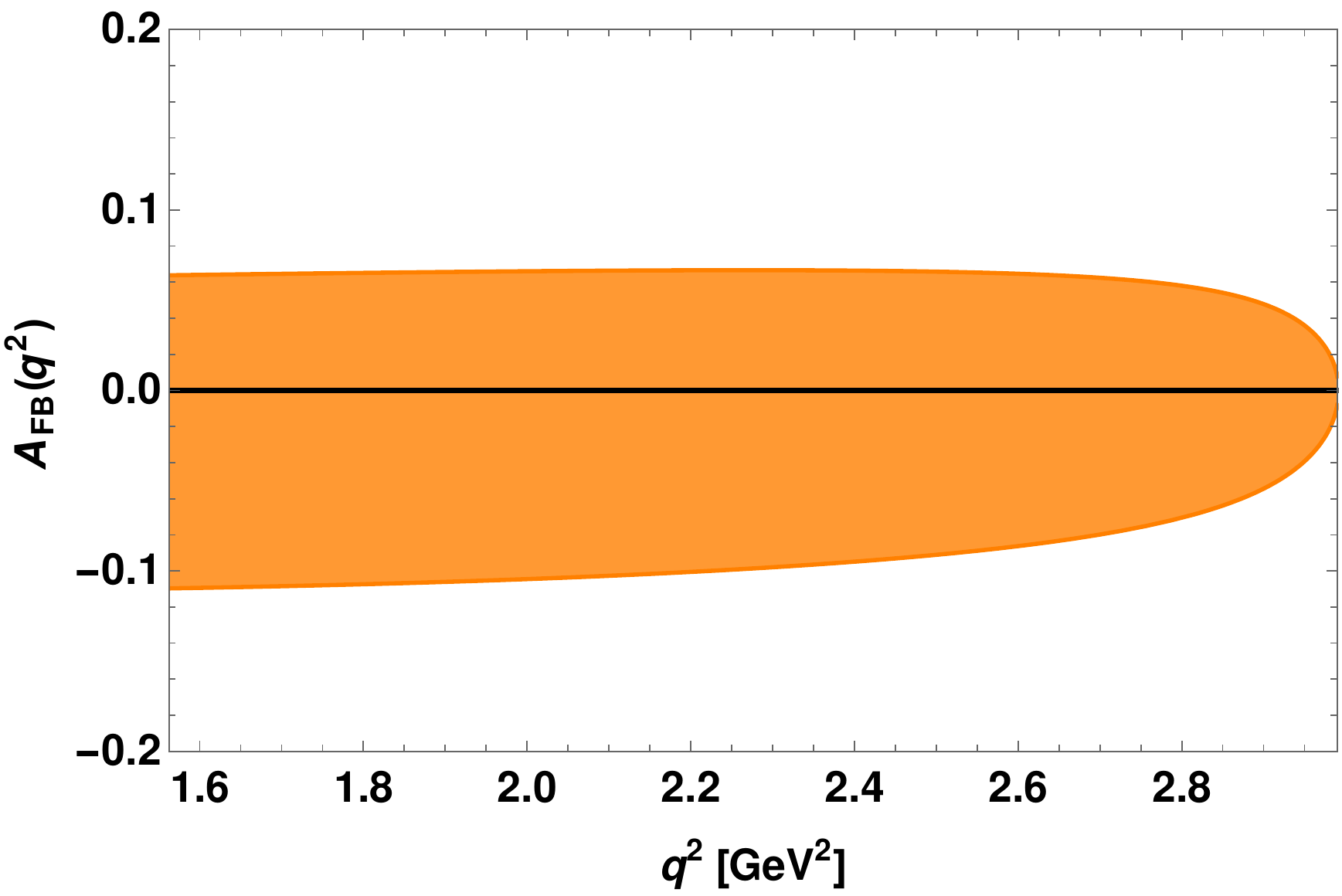}
\caption{The variation of forward-backward asymmetry of $ D^+ \rightarrow \pi^+ \mu^+ \mu^-$  with respect to low $q^2$ (left panel) and high $q^2$ (right panel) in   scalar leptoquark model.}
\end{figure}
\begin{figure}[h]
\centering
\includegraphics[width=7.0cm,height=5.5cm]{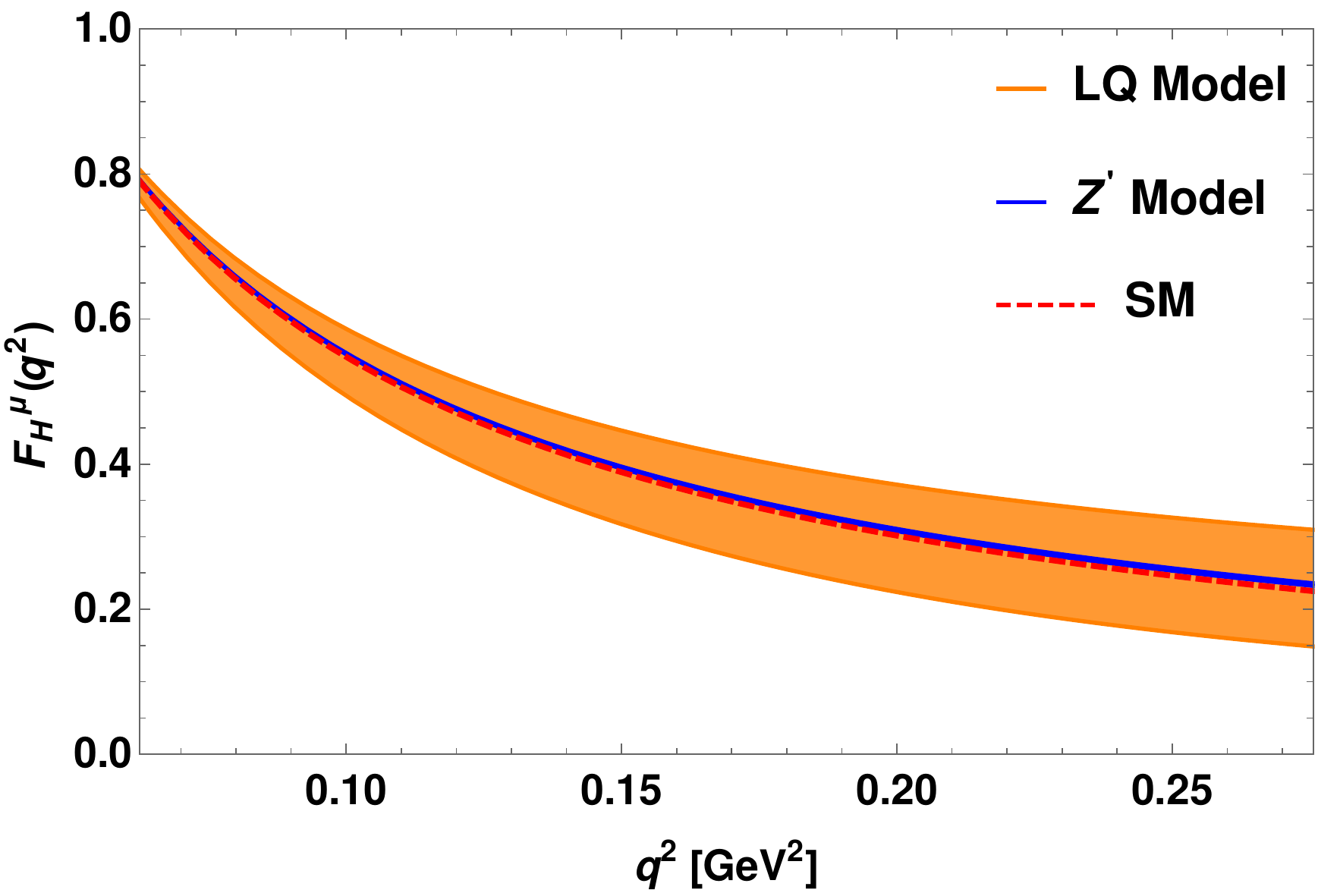}
\hspace{0.25 true cm}
\includegraphics[width=7.0cm,height=5.5cm]{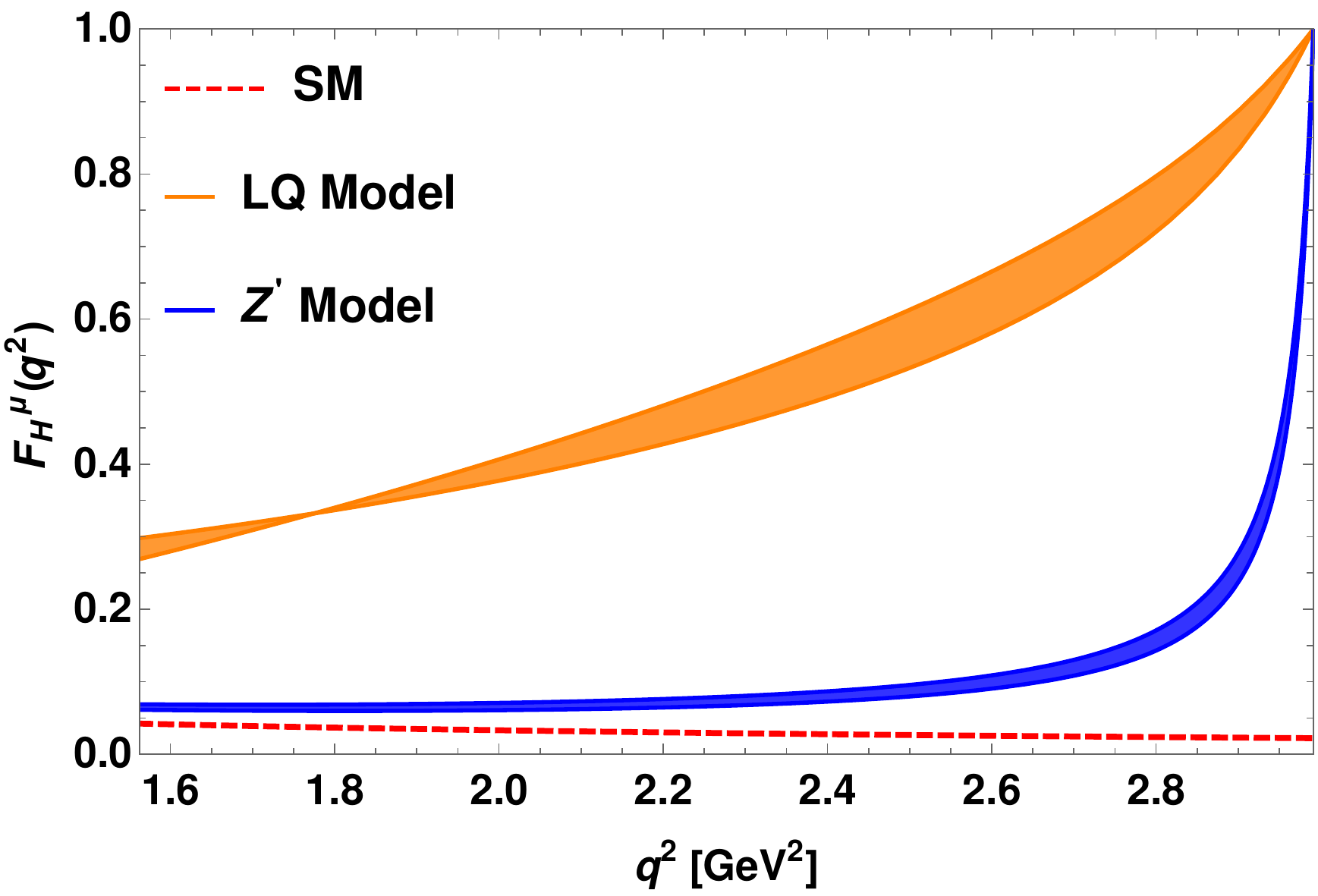}
\caption{The variation of flat term of $ D^+ \rightarrow \pi^+ \mu^+ \mu^-$  with respect to low $q^2$ (left panel) and high $q^2$ (right panel) in   scalar leptoquark  and $Z^\prime$ models.}
\end{figure}
\begin{table}[h]
\caption{The predicted branching ratios for $D^{+(0)} \to \pi^{+(0)}  \mu^+ \mu^-$  processes in both the low  $q^2$ and high $q^2$ region in the scalar $X(3,2,7/6)$ LQ and $Z^\prime$  model. This also contains the resonant and nonresonant SM branching ratios. }
\begin{center}
\begin{tabular}{| c | c | c | c |}
\hline
 & Decay process & $D^+ \to \pi^+ \mu^+ \mu^-$  & $D^0 \to \pi^0 \mu^+ \mu^-$  \\
\hline
 \hline
  & Nonresonant SM &   $(3.02 \pm 0.483) \times 10^{-13}$ &$ (1.19 \pm 0.19) \times 10^{-13}$  \\
  & Resonant SM &   $(1.36 - 2.4) \times 10^{-10}$ &$ (5.66-9.89) \times 10^{-11}$  \\

~low $q^2$~ & LQ model & $(2.6-8.68) \times 10^{-10}$ &  $(1.02-3.4) \times 10^{-10}$\\
& $Z^\prime$ model & $~(0.65-1.18) \times 10^{-12}$~ &~$(2.55-4.62) \times 10^{-13}$ ~\\

 &~Expt. limit ($90\%$ CL) ~& $2 \times 10^{-8}$ \cite{LHCb-exp} & $\cdots$ \\
 
 \hline
 
 & Nonresonant SM &   $(5.14\pm 0.82) \times 10^{-13}$ &$(2 \pm 0.32) \times 10^{-13}$  \\
  & Resonant SM &   $(1.25 - 3.29) \times 10^{-10}$ &$ (0.456-1.24) \times 10^{-10}$  \\
high $q^2$ &LQ model & $(1.32-3.36) \times 10^{-9}$ &  $(0.513-1.3)\times 10^{-9}$\\

 &$Z^\prime$ model & ~$(1.4-2.78) \times 10^{-12}$~ &~$(0.545-1.08) \times 10^{-12}$ ~\\
 &~Expt. limit ($90\%$ CL) ~& $2.6 \times 10^{-8}$ \cite{LHCb-exp}& $\cdots$ \\
 \hline
\end{tabular}
\end{center}
\end{table} 
\section{$D^{+ (0)} \to \pi^{+ (0)} \mu^- e^+$ }
Since the individual lepton flavour number is conserved in the standard model, the observation of lepton flavour violation  in the near  future will provide unambiguous signal of new physics beyond the SM. The observation of neutrino oscillation implies the violation of lepton flavour in neutral sector and it is expected that there could be  FCNC transitions in the charged  lepton sector as well, such as  $l_i \to l_j \gamma$, 
$l_i \to l_j l_k \bar{l}_k$, $B \to l_i^\mp l_j^\pm$ and $B \to K^{(*)} l_i^\mp l_j^\pm$ etc. The LFV decay modes proceed   through box diagrams with tiny neutrino masses in the loop, thus become very rare in the SM. However, these modes can  occur at tree level in the leptoquark and $Z^\prime$ models, thus can provide observable signature in the high luminosity  experiments.  In this section, we would like to study the lepton flavour violating semileptonic decay  process $D^+ \to \pi^+ \mu^- e^+$. Due to the absence of intermediate states, these LFV processes have no long distance
QCD contributions and dominant $\phi$, $\omega$  resonance backgrounds. The general expression for the transition amplitude of $D^+ \to \pi^+ \mu^- e^+$ process in a generalized new physics model, is given by 
\begin{equation}
\mathcal{M} = -\frac{G_F \alpha_{em} \lambda_b}{\sqrt{2}\pi} f_+ (q^2) \Bigg[F_S (\bar{\mu}e) + F_P (\bar{\mu} \gamma_5 e) + F_V p_D^\mu \left(\bar{\mu} \gamma_\mu e \right) + F_A p_D^\mu \left(\bar{\mu} \gamma_\mu 
\gamma_5 e\right) \Bigg]\;, \label{amp}
\end{equation}
where the functions $F_i$, $i=V, A, S, P$ are defined as
\bea
F_V &=& K_9^{NP}+K_9^{\prime NP}\;, \hspace{3cm} F_A = K_{10}^{NP}+K_{10}^{\prime NP}\;, \nn\\
F_S &= & \frac{1}{2} (K_S^{NP}+K_S^{\prime NP}) \frac{M_D^2-M_\pi^2}{m_c}\frac{f_0 (q^2)}{f_+ (q^2)} \nn \\ &+&  \frac{1}{2} \left(K_9^{NP}+K_9^{\prime NP}\right) (m_e - m_\mu) \Bigg[\frac{M_D^2 - M_\pi ^2}{q^2}  \left( \frac{f_0 (q^2)}{f_+ (q^2)} - 1 \right) -1 \Bigg]  \;,\nn\\
F_P & = & \frac{1}{2} (K_P^{NP}+K_P^{\prime NP}) \frac{M_D^2-M_\pi^2}{m_c}\frac{f_0 (q^2)}{f_+ (q^2)}\nn \\ &+&  \frac{1}{2}\left( K_{10}^{NP}+K_{10}^{\prime NP}\right) (m_\mu + m_e)\left[\frac{M_D^2 - M_\pi^2}{q^2} \left( \frac{f_0 (q^2)}{f_+ (q^2)} - 1 \right)-1 \right] .
\eea
Here the Wilson coefficients $(K_i^{NP})$ involve the combination of LQ couplings as $Y_{\mu c}^L Y_{e u}^{L/R*}$ instead of $Y_{\mu c}^L Y_{\mu u}^{L/R*}$ in Eqn. (8). 
Now using Eqn. (\ref{amp}), the differential decay distribution for the  $D^+ \to \pi^+ \mu^- e^+$ process with respect to $q^2$ and $\cos\theta$ ($\theta$ is the angle between the $D$ and  $\mu^-$   in the $\mu-e$ rest frame) is given as
\begin{equation}
\frac{d^2\Gamma}{dq^2 d\cos\theta} = A_l (q^2) + B_l (q^2) \cos\theta + C_l (q^2) \cos^2\theta\;,
\end{equation}
where
\bea
A_l (q^2) & =& 2\Gamma_0 \frac{\sqrt{\lambda_1 \lambda_2}}{q^2} f_+(q^2)^2 \Bigg[ \frac{\lambda_1}{4}\left(|F_V|^2 + |F_A|^2 \right) 
 + |F_S|^2 \left (q^2 - (m_\mu + m_e)^2\right )\nn\\ 
& + & |F_P|^2 \left (q^2 - (m_\mu - m_e)^2 \right )  + |F_A|^2 M_D^2 (m_\mu + m_e)^2 + |F_V|^2 M_D^2 (m_\mu - m_e)^2\nn  \\
& + & \left(M_D^2 - M_\pi^2 + q^2 \right) \Big( (m_\mu + m_e)  {\rm Re} (F_P F_A^*) + (m_e - m_\mu) {\rm Re} (F_S F_V^*)\Big) \Bigg]\;,
\eea
\begin{equation}
B_l (q^2) = 2\Gamma_0 \frac{\sqrt{\lambda_1 \lambda_2}}{q^2} f_+(q^2)^2 \Bigg[ (m_\mu + m_e) {\rm Re} (F_S F_V^*) + (m_e - m_\mu) {\rm Re} (F_P F_A^*) \Bigg]\;, \hspace{1.2cm}
\end{equation}
\begin{equation}
C_l (q^2) = - 2\Gamma_0  f_+(q^2)^2 \frac{(\lambda_1 \lambda_2)^{3/2}}{4 q^6} \left(|F_A|^2 + |F_V|^2 \right)\;, \hspace{6cm}
\end{equation}
and
\bea
 \lambda_1 = \lambda (M_D^2, M_\pi^2, q^2),  \hspace{1cm} \lambda_2 = \lambda (q^2,m_\mu^2, m_e^2). 
\eea
For numerical estimation in the leptoquark model, we use the constrained leptoquark couplings obtained  from $D^0 \to \mu^+ \mu^-$ process  and assume that the coupling between different generation of quarks and leptons follow the
simple scaling laws, i.e. $Y_{ij}^{L(R)} / Y_{ii}^{L(R)}  = (m_i/m_j)^{1/2}$ with $j > i$. As discussed in \cite{ansatz1, ansatz2}, such pattern of ansatz can explain the decay widths of radiative LFV decay $\mu \to e \gamma$.   
Now using such ansatz, the variation of branching ratio with respect to $q^2$ for $D^{+(0)} \to \pi^{+(0)} \mu^- e^+$  process in the leptoquark model is shown in left panel of Fig. 6 and the corresponding integrated value  is given in Table III.  In this mode, the forward backward asymmetry depends on $K_{9, 10}^{(')NP}$ Wilson coefficients which give nonzero contribution.   The left panel of Fig. 7 shows  the $q^2$ variation of the forward backward asymmetry and the corresponding integrated value is found to be $(0.039 \to 0.047)$. The  variation of the flat term with  respect to $q^2$ is presented in the left panel of Fig. 8 and the integrated value is $(0.137\to 0.33)$. 

For the $Z'$ model, we consider the constraint on the coupling of $Z^\prime$ boson to the quarks, obtained from the $D^0-\bar{D}^0$ mixing and $D^0 \to \mu^+ \mu^-$ process as given in Eqn. (\ref{D0-Zprime}) and (\ref{D0mu-Zprime}). For the lepton flavour violating coupling,  the constraint is taken from $\mu^- \to e^- e^+ e^-$ process, as discussed   in section IV. Thus, using Eqn (\ref{D0-Zprime}), (\ref{D0mu-Zprime}) and (\ref{D0mueee-Zprime}), the predicted branching ratio  of $D^{+(0)} \to \pi^{+(0)} \mu^- e^+$  process in the $Z^\prime$ model is given in Table III and the $q^2$ variation  of $D^+ \to \pi^+ \mu^- e^+$ process  is shown in  Fig. 6 (right panel). The forward-backward asymmetry variation is shown in right panel of  Fig. 7 and the predicted value is $-1.15 \times 10^{-3}$, which is very small. In Fig. 8 (right panel), we show the plot for $q^2$ variation of the  flat term and the integrated value is $0.158$.

From Table III, one can note that the predicted branching ratios are well below the present experimental limit for the $D^{+} \to \pi^{+} \mu^- e^+$  process. Although  there is no experimental bound on $D^{0} \to \pi^{0} \mu^- e^+$  process so far,  the experimental upper limit on branching ratios of  $D^{0} \to \pi^{0} \mu^\mp e^\pm$  process is known, which is given as ${\rm BR}(D^{0} \to \pi^{0} \mu^\mp e^\pm)= {\rm BR}(D^{0} \to \pi^{0} \mu^- e^+ + \pi^0 \mu^+ e^-) < 8.6 \times 10^{-5}$. Our results   for $D^{0} \to \pi^{0} \mu^- e^+$ process in both the leptoquark and $Z^\prime$ models are found to be within the above experimental bound. 
\begin{figure}[h]
\centering
\includegraphics[width=7.0cm,height=5.5cm]{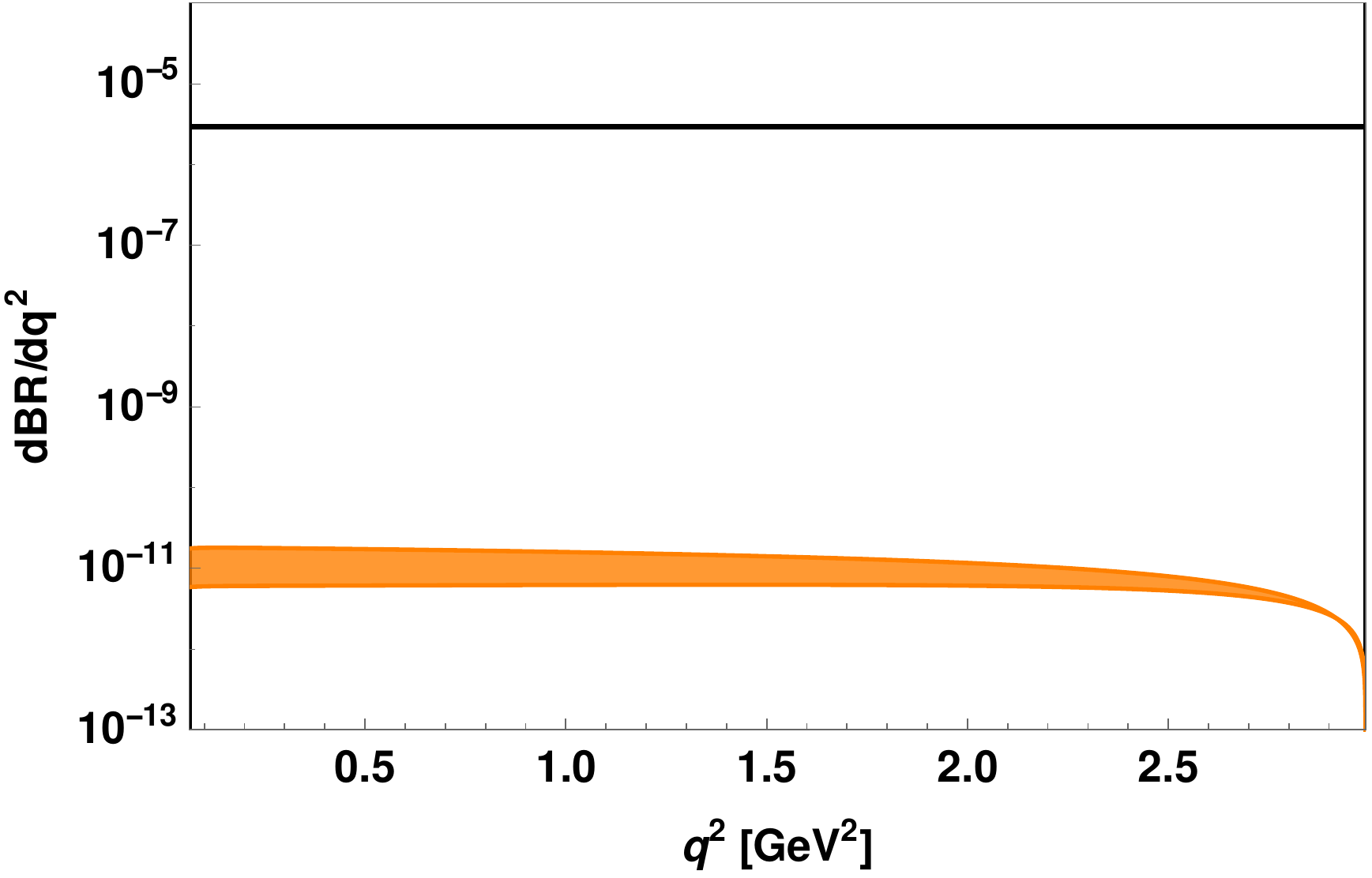}
\hspace{0.25 true cm}
\includegraphics [width=7.0cm,height=5.5cm]{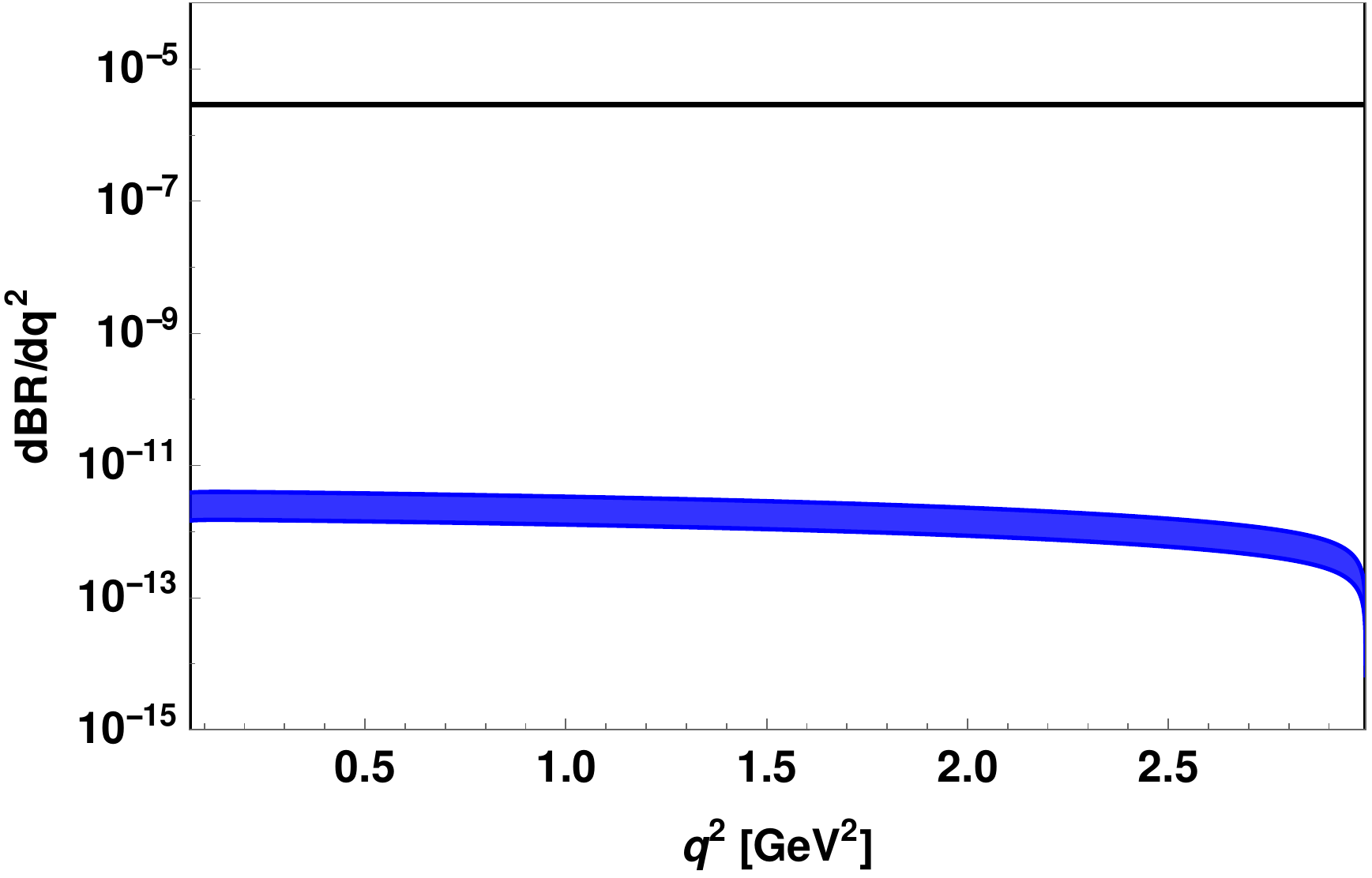}
\caption{The variation of branching ratio of LFV $ D^+ \rightarrow \pi^+ \mu^- e^+$ process in the leptoquark model (left panel) and generic $Z^\prime$ model (right panel)  with respect to  $q^2$. The solid black lines represent the $90\%$ CL experimental upper bound \cite{pdg}. }
\end{figure}
\begin{figure}[h]
\centering
\includegraphics[width=7.0cm,height=5.5cm]{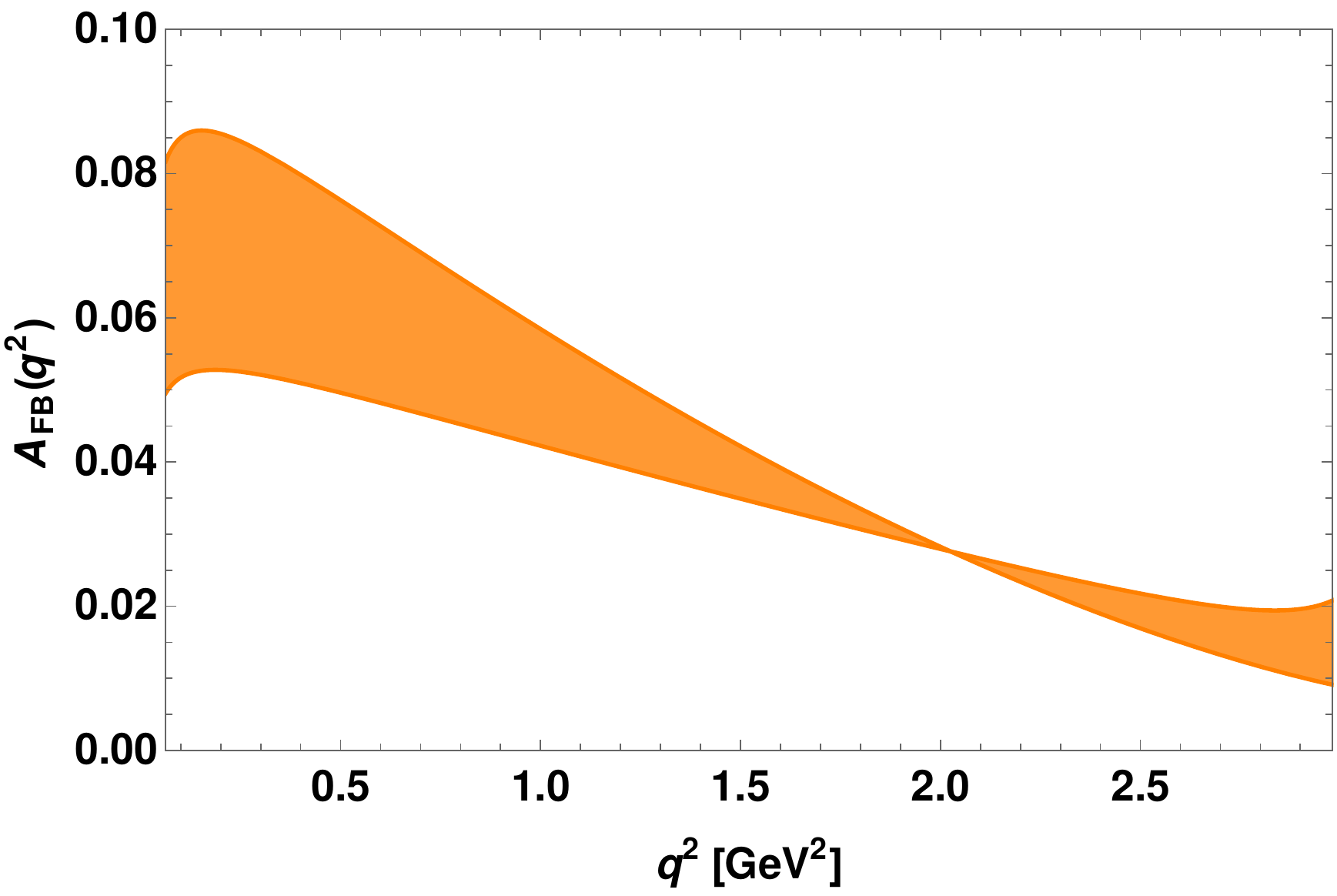}
\hspace{0.25 true cm}
\includegraphics [width=7.0cm,height=5.5cm]{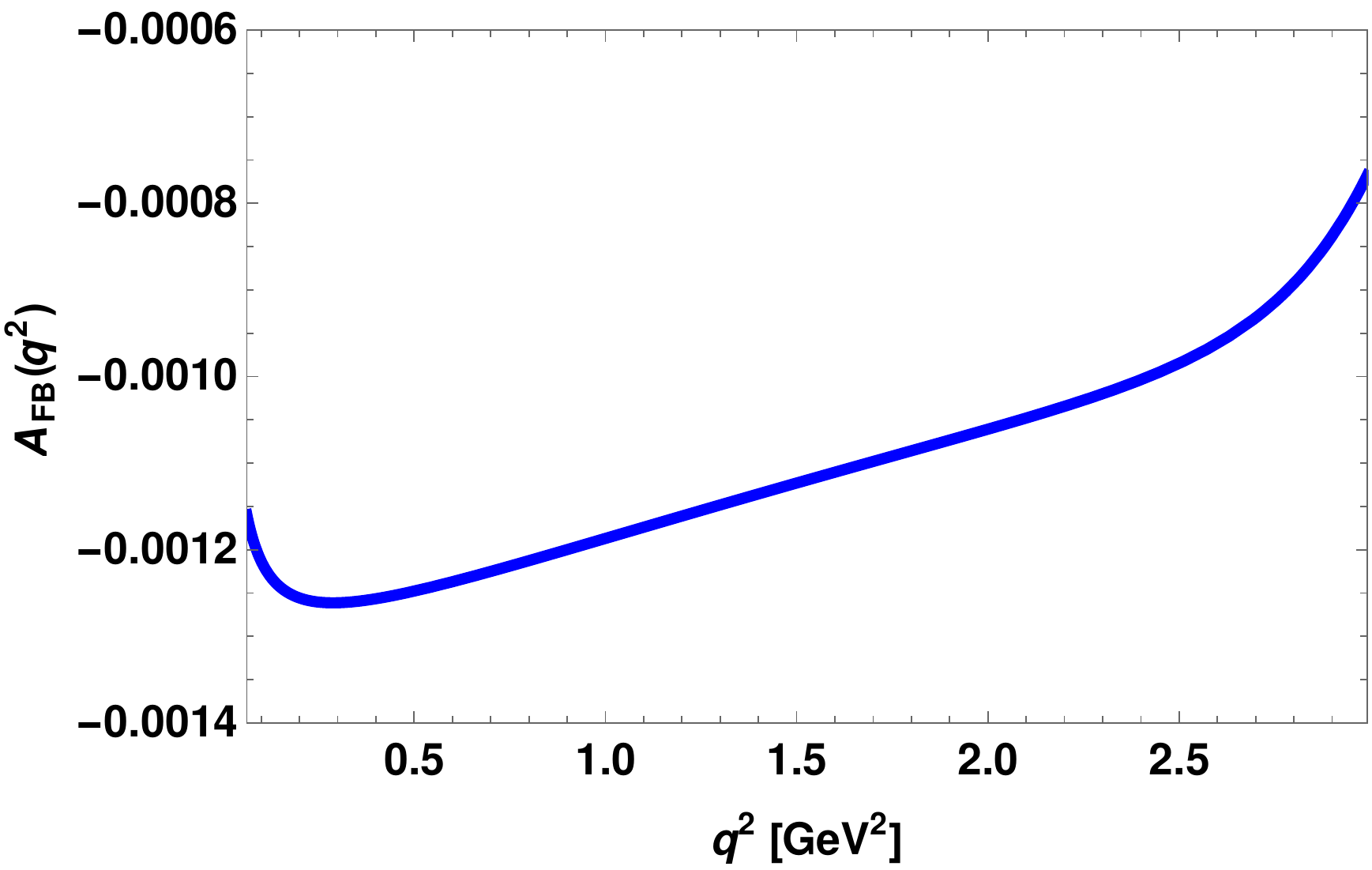}
\caption{The variation of forward-backward asymmetry of LFV $ D^+ \rightarrow \pi^+ \mu^- e^+$ process in the leptoquark model (left panel) and generic $Z^\prime$ model (right panel)  with respect to  $q^2$. }
\end{figure}
\begin{figure}[h]
\centering
\includegraphics[width=7.0cm,height=5.5cm]{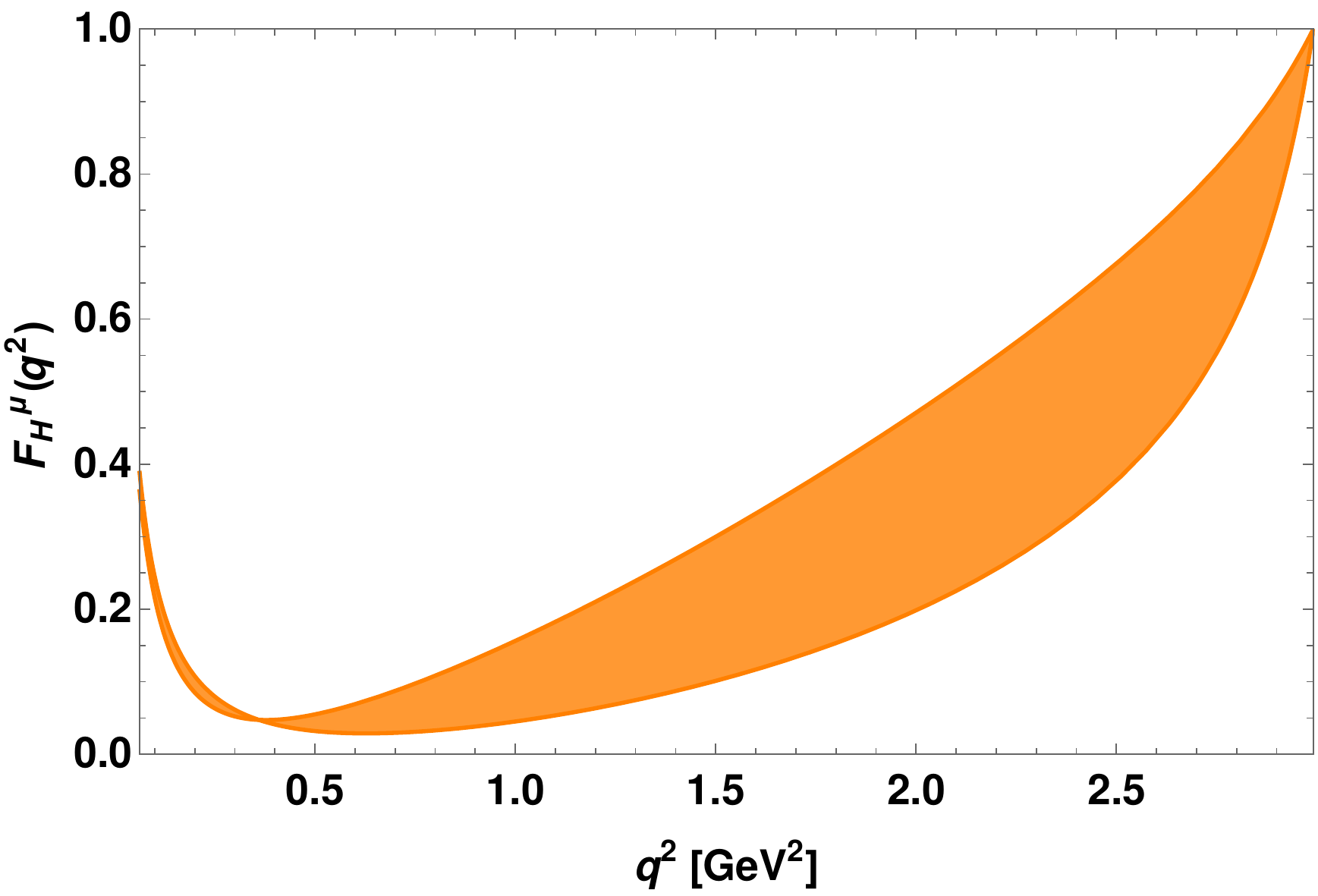}
\hspace{0.25 true cm}
\includegraphics [width=7.0cm,height=5.5cm]{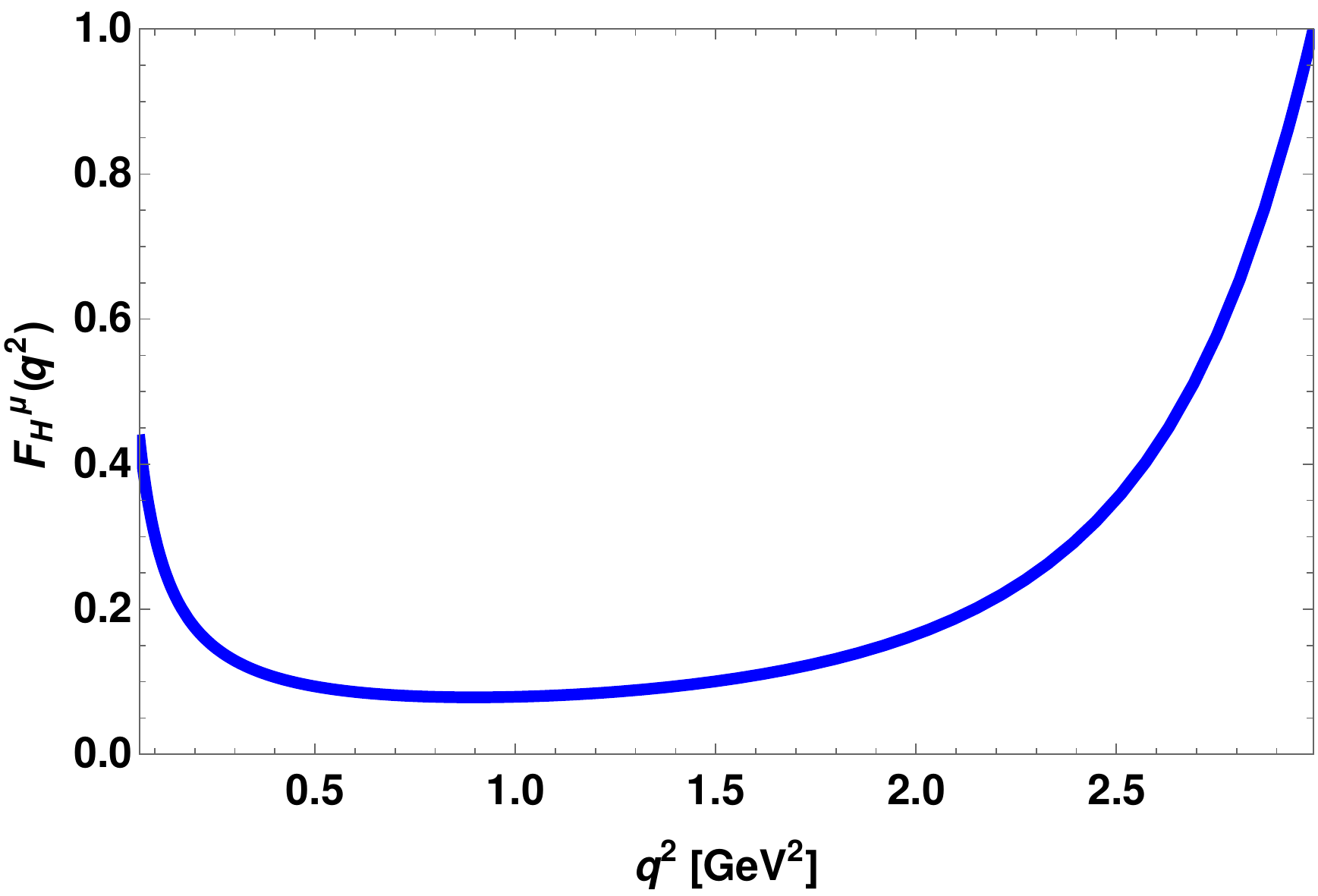}
\caption{The variation of flat term of   LFV $ D^+ \rightarrow \pi^+ \mu^- e^+$ process in the leptoquark model (left panel) and generic $Z^\prime$ model (right panel)  with respect to  $q^2$.}
\end{figure}

\begin{table}[h]
\caption{The predicted branching ratios for $D^{+(0)} \to \pi^{+(0)}  \mu^- e^+$ lepton flavour violating processes in the scalar $X(3,2,7/6)$ LQ and $Z^\prime$  model. The  present upper limit on the branching ratio   ${\rm BR}(D^{0} \to \pi^{0} \mu^\mp e^\pm)= {\rm BR}(D^{0} \to \pi^{0} \mu^- e^+ + \pi^0 \mu^+ e^-) < 8.6 \times 10^{-5}$ \cite{pdg}.}
\begin{center}
\begin{tabular}{| c | c | c | }
\hline
 Decay process & $D^+ \to \pi^+ \mu^- e^+$ &$D^0 \to \pi^0 \mu^- e^+$    \\
\hline
LQ model  &   $ (1.67-3.72) \times 10^{-11}$ & $ (0.56-1.4) \times 10^{-11}$   \\
$Z^\prime$ model  & $(2.95-7.8)\times 10^{-12}$ & $(1.15-3.04) \times 10^{-12}$\\
~ Experimental limit ~&$< 2.9 \times 10^{-6}$ \cite{pdg}  & $\cdots$\\
 \hline
\end{tabular}
\end{center}
\end{table}
\section{$D^0 \to \mu^- e^+ (\tau^- e^+)$ LFV decay process}
 Recently LHCb put the upper limit on branching ratio of the   $D^0 \to \mu^\mp e^\pm $ lepton flavour violating decay  mode  as \cite{D0exp}
\bea
\rm{BR}(D^0 \to \mu^\mp e^\pm) \simeq \rm{BR}(D^0 \to \mu^- e^+ + \mu^+ e^-) < 1.3 \times 10^{-8}.
\eea
Neglecting the mass of electron, the branching ratio of $D^0\to \mu^- e^+$ process is given by  \cite{Hiller}

\begin{align}
 {\rm BR}(D^0\to \mu^- e^+)&= \tau_{D} \frac{G_F^2\alpha_e^2 M_{D}^5f_{D}^2 |\lambda_b|^2}{64\pi^3}  \left(1-\frac{m_\mu^2}{M_{D}^2}\right)^2  \bigg[\left|\frac{K_S^{NP}-K_S^{\prime NP}}{m_c} + \frac{ m_\mu }{M_{D}^2} \left(K_9^{NP}-K_9^{\prime NP}\right) \right|^2\nonumber\\
 &+\left|\frac{K_P^{NP}-K_P^{ \prime NP}}{m_c}+\frac{m_\mu}{M_{D}^2}\left(K_{10}^{NP}-K_{10}^{\prime NP}\right)\right|^2\bigg] \, .       \label{eq:B_D0mue} 
\end{align}
We use the  scaling ansatz as discussed in the previous section to compute the required leptoquark coupling for  $D^0\to \mu^- e^+$ process and the predicted branching ratio is found to be 
\bea
\rm{BR}(D^0 \to \mu^- e^+) = (3.18-4.8) \times 10^{-11}.
\eea
Now using Eqns. (\ref{D0-Zprime}), (\ref{D0mu-Zprime}) and (\ref{D0mueee-Zprime}), the predicted branching ratio of this LFV process in the $Z^\prime$ model  is 
\bea
\rm{BR}(D^0 \to \mu^- e^+) \simeq 6.1 \times 10^{-17}.
\eea
The predicted branching ratio is although small, but  can be searched at LHCb experiment. The   exploration/observation  of  this decay 
mode would definitely shed some light in the leptoquark scenarios.

Similarly using the new Wilson coefficient generated via  leptoquark exchange, the  branching ratio for $D^0\to \tau^- e^+$ process is
found to be
\bea
\rm{BR}(D^0 \to \tau^- e^+) =  (2.84-9.75) \times 10^{-14}.
\eea 
However, there is no experimental  observation of  lepton flavour violating $D^0\to \tau^- e^+$ process.  The constraint on $Z^\prime$ coupling  to tau and electron is obtained  from the $\tau^- \to e^- e^+ e^-$ process. Using  Eqn. (\ref{D0taueee-Zprime}), the  branching ratio for $D^0\to \tau^- e^+$ process  in $Z^\prime$ model is  given as
\bea
\rm{BR}(D^0 \to \tau^- e^+) =  (0.73-1.94) \times 10^{-15}.
\eea
So far there is no experimental evidence on the LFV $D^0 \to \tau^- e^+$ decay process.   Our results for $D^0 \to \tau^- (\mu^-) e^+$ process is comparable with \cite{Hiller, Pakvasa2, Hiller3}. 


\section{conclusion}
In this paper we have studied the rare decays of $D$ meson in both scalar leptoquark and generic $Z^\prime$ models. We have considered the  simple renormalizable scalar leptoquark model  with the requirement that proton decay would not be induced in perturbation theory. The leptoquark parameter space is constrained using the  present upper limit on branching ratio of $D^0 \to \mu^+ \mu^-$ process and the  $D^0-\bar{D}^0$ oscillation data.  For the   $Z^\prime$ model, the constraints on  $Z^\prime$ couplings are obtained from the mass difference of $D^0-\bar{D}^0$ mixing,  $D^0 \to \mu^+ \mu^-$ process and the lepton flavour violating $\tau^-(\mu^-) \to e^- e^+ e^-$ processes.  Using the constrained parameter space, we estimated the branching ratios, forward backward asymmetry parameters and the flat terms  in $D^{+(0)} \to \pi^{+(0)} \mu^+ \mu^-$ processes. The branching ratios in the LQ model are
found to be $\sim{\cal O}(10^{-10})$, which are larger than the corresponding SM predictions in  the very low and very high $q^2$ regimes. If these branching ratios will be observed in near future they would provide indirect hints of leptoquark
signal.  Furthermore we estimated the branching ratios of lepton flavour violating $D^{+(0)} \to \pi^{+(0)} \mu^- e^+$ and $D^0 \to \mu(\tau)^- e^+ $ processes, which are  
found to be rather small.  We also estimated the forward-backward asymmetry parameter and the flat term for the LFV decays.

{\bf Acknowledgments}

We  would like to thank  Science and Engineering Research Board (SERB), Government of India for financial support through grant No. SB/S2/HEP-017/2013.    

\end{document}